\documentclass[a4paper,fleqn,usenatbib]{mnras}
\usepackage{ragged2e} 
\usepackage[T1]{fontenc}
\usepackage{ae,aecompl}
\usepackage{graphicx}	
\usepackage{amsmath}	
\usepackage{amssymb}	
\usepackage{cleveref}   
\usepackage{hyperref}   
\usepackage[countmax]{subfloat}
\usepackage{booktabs}
\newcommand{\RNum}[1]{\uppercase\expandafter{\romannumeral #1\relax}}

\title[]{Study of correlation between optical flux and polarization variations in BL Lac objects}

\author[Bhoomika et al.]{
Bhoomika Rajput$^{1,2}$\thanks{E-mail: bhoomikarjpt2@gmail.com},
Ashwani Pandey$^{2}$,
C. S. Stalin$^{2}$,
and Blesson Mathew$^{1}$
\\
$^{1}$Department of Physics, CHRIST (Deemed to be University), Hosur Road, Bangalore 560 029, India\\
$^{2}$Indian Institute of Astrophysics, Block II, Koramangala, Bangalore 560034, India
}
\date{Accepted XXX. Received YYY; in original form ZZZ}

\pubyear{2020}

\begin{document}
\label{firstpage}
\pagerange{\pageref{firstpage}--\pageref{lastpage}}
\maketitle
\begin{abstract}
Polarized radiation from blazars is one key piece of evidence for synchrotron radiation at low energy, which also shows variations. We present here our results on the correlation analysis between optical flux and polarization degree (PD) variations in a sample of 11 BL Lac objects using $\sim$ 10 years of data from the Steward Observatory. We carried out the analysis on long-term ($\sim$ several months) as well as on short-term timescales ($\sim$ several days). On long-term timescales, for about 85\% of the observing cycles, we found no correlation between optical flux and PD. On short-term timescales, we found a total of 58 epochs with a significant correlation between optical flux and PD, where both positive and negative correlation were observed. In addition, we also found a significant correlation between optical flux and $\gamma$-ray flux variations on long-term timescales in 11\% of the observing cycles. The observed PD variations in our study cannot be explained by changes in the power-law spectral index of the relativistic electrons in the jets. The shock-in-jet scenario is favoured for the correlation between optical flux and PD, whereas the anti-correlation can be explained by the presence of multi-zone emission regions. The varying correlated behaviour can also be explained by the enhanced optical flux caused by the newly developed radio knots in the jets and their magnetic field alignment with the large scale jet magnetic field.
\end{abstract}



\section{INTRODUCTION} \label{sec:intro}
Blazars are the jetted class of active galactic nuclei (AGN), and they are among the most luminous (10$^{42}$-10$^{48}$ erg s$^{-1}$) objects in the extragalactic $\gamma$-ray sky. The accretion of matter onto the super massive black holes is thought to be the source of immense energy of blazars. The mass of super massive black holes in blazars ranges from 10$^{6}$-10$^{10}$ M$_{\odot}$ \citep{1969Natur.223..690L,1973A&A....24..337S}. Blazars have relativistic jets that are oriented close to the line of sight to the observer \citep{1995PASP..107..803U}, and as a result the radiation from their jets is highly Doppler boosted. They emit radiation over the entire accessible electromagnetic spectrum (i.e. from radio to $\gamma$-rays) and their emission is also known to be highly variable \citep{1996ASPC..110..391U}. In addition to flux variations, blazars are also known to show large optical polarization and polarization variability \citep[and references therein]{2019Galax...7...85Z}.\\
Blazars are divided into two subclasses based on the emission lines in their optical spectra. These subclasses are flat spectrum radio quasars (FSRQs) with strong emission lines (EW $>$ 5\AA) and BL Lac objects with no or weak emission lines (EW $<$ 5\AA) \citep{1991ApJS...76..813S}. A more physical criterion based on the accretion ratio was introduced by \cite{2011MNRAS.414.2674G} and \cite{2012MNRAS.421.1764S} to distinguish between these two subclasses. According to this scheme, FSRQs have the ratio of the luminosity of the broad line region $L_{BLR}$ to the Eddington luminosity $>$ 5 $\times$ $10^{-4}$ while it is less than 5 $\times$ $10^{-4}$ for BL Lacs.\\
The broad band spectral energy distribution (SED) of blazars consists of two hump structure, in which the lower energy hump is well understood by the synchrotron emission process of the relativistic electrons in the jet of these sources and the high energy hump is attributed to the inverse Compton (IC) process \citep{2010ApJ...716...30A}. The seed photons which take part in the IC process can be present either within the jet (synchrotron-self Compton; SSC; \citealt{1985ApJ...298..114M,1989ApJ...340..181G} or outside of the jet such as BLR, torus and accretion disk (external Compton; EC; \citealt{1987ApJ...322..650B}). FSRQs and BL Lac sources are further divided based on the position of synchrotron peak frequency ($\nu^{syn}_{peak}$) in the broad band SED \citep{2010ApJ...716...30A}. These are low-synchrotron peaked blazars (LSP; $\nu^{syn}_{peak}$ $<$ $10^{14}$ Hz), intermediate-synchrotron peaked blazars (ISP; $10^{14}$ Hz $<$ $\nu^{syn}_{peak}$ $<$ $10^{15}$ Hz) and high-synchrotron peaked blazars (HSP; $10^{15}$ $<$ $\nu^{syn}_{peak}$ $<$ $10^{17}$ Hz) \citep{2011ApJ...743..171A, 2015ApJ...810...14A}. FSRQs are mostly LSP blazars, while BL Lac sources exhibit all three blazar behaviours. Extreme nature is also present in some BL Lacs, classified as extremely high-synchrotron peaked BL Lacs (EHBLs or extreme blazars; $\nu^{syn}_{peak}$ $>$ $10^{17}$ Hz) \citep{2019MNRAS.486.1741F}. \\ 
 Blazars show highly variable polarization behaviour in the optical band \citep{1998AJ....116.2119V, 2017ApJS..232....7F}. The observation of polarization features from blazar jets provides vital information for studying the structural information of the magnetic field within the jets. The order of the magnetic field vectors or the electron energy distribution within the emission region could be represented by the degree of optical polarization, whereas the electric vector position angle (EVPA) could be related to the direction of the magnetic field vector along the line of sight \citep{1980ARA&A..18..321A, 2000ApJ...541...66L}. The study of correlation between optical flux and polarization variations can provide important insights into how polarization behaves in different optical flux states. Studies available in literature on correlation between optical flux and polarization variations show( i) positive correlation between optical flux and polarization variations \citep{2008ApJ...672...40H, 2013ApJS..206...11S} (ii) negative correlation between optical flux and PD \citep{1991AJ....101.2017S, 2015ApJ...809L..27B, 2017ApJ...835..275R}, (iii) varied behaviours between optical flux and PD variations that includes positive, negative and no correlation \citep{2002A&A...385...55H, 2006ChJAS...6a.247J, 2014ApJ...781L...4G, 2019MNRAS.486.1781R, 2020MNRAS.492.1295P, 2022MNRAS.510.1809P}.\\
Furthermore, the nature of the $\gamma$-ray emission process can be constrained by analysis of the correlation between $\gamma$-ray flux and optical polarization position angle. The observation obtained from $\it{Fermi}$ Gamma-ray Space Telescope (hereinafter $\it{Fermi}$) reported the various behaviour of correlation between $\gamma$-ray flux and optical flux \citep{2009ApJ...697L..81B, 2012ApJ...749..191C, 2014ApJ...797..137C, 2015ApJ...804..111M, 2019MNRAS.486.1781R, 2020MNRAS.498.5128R, 2021ApJ...906....5A, 2021MNRAS.504.1772R} and between $\gamma$-ray flux and optical polarization position angle \citep{2010Natur.463..919A, 2010ApJ...710L.126M, 2015MNRAS.453.1669B, 2018MNRAS.474.1296B}. \\
However, these findings show that the polarization degree and position angle, and their correlation with different energy bands, have a complex behaviour. So, in order to look for anomalous differences in optical flux and polarization behaviour in blazars, we aim for a systematic investigation of the correlation between optical flux and polarization variations in blazars. Towards this goal, we selected a sample of 8 FSRQs and 11 BL Lac sources from the archives of Steward Observatory. The results of the 8 FSRQ blazars were reported by \cite{2022MNRAS.510.1809P}. In this work, we present results on the systematic investigation of the correlation between optical V-band flux and polarization variations in 11 BL Lac sources from 2008 to 2018 on long-term ($\sim$few months) as well as short-term ($\sim$few days) timescales. On short-term timescales, we also investigated the correlation behaviour between optical V-band flux and $\gamma$-ray flux. The details of the 11 BL Lac sources are given in Table \ref{tab:src_list}. In Section \ref{sec:data}, we give the details of the data used in this work. In Section \ref{sec:lightcurve}, the multiwavelength light curves of the BL Lac sources are described. In Section \ref{sec:analysis}, the details of the analysis are given. The results and the discussion are given in Section \ref{sec:results} and Section \ref{sec:diss}, respectively. We give the summary of the work in Section \ref{sec:sum}.
 
 \begin{table}
\centering
\caption{\label{tab:src_list}Details of the BL Lac sources.}
\begin{tabular} {lcccr} \hline
Source  &   RA ($\alpha_{\rm 2000}$)  &  Dec ($\delta_{\rm 2000}$)  & Redshift (z)  & \\ \hline
3C 66A           & $02^h22^m40^s$ & $+43^{\circ}02^{\prime}08^{\prime\prime}$ & 0.340 & \\
AO 0235$+$164    & $02^h38^m39^s$ & $+16^{\circ}36^{\prime}59^{\prime\prime}$ & 0.940 & \\
S5 0716$+$714    & $07^h21^m53^s$ & $+71^{\circ}20^{\prime}36^{\prime\prime}$ & 0.310 & \\
OJ 287           & $08^h54^m49^s$ & $+20^{\circ}06^{\prime}31^{\prime\prime}$ & 0.306 & \\
MRK 421          & $11^h04^m27^s$ & $+38^{\circ}12^{\prime}32^{\prime\prime}$ & 0.031 & \\
W Comae          & $12^h21^m32^s$ & $+28^{\circ}13^{\prime}59^{\prime\prime}$ & 0.103 & \\
MRK 501          & $16^h53^m52^s$ & $+39^{\circ}45^{\prime}37^{\prime\prime}$ & 0.034 & \\
1ES 1959$+$650   & $19^h59^m60^s$ & $+65^{\circ}08^{\prime}55^{\prime\prime}$ & 0.048 & \\
PKS 2155$-$304   & $21^h58^m52^s$ & $-30^{\circ}13^{\prime}32^{\prime\prime}$ & 0.116 & \\
BL Lac           & $22^h02^m43^s$ & $+42^{\circ}16^{\prime}40^{\prime\prime}$ & 0.069 & \\
1ES 2344$+$514   & $23^h47^m05^s$ & $+51^{\circ}42^{\prime}18^{\prime\prime}$ & 0.044 & \\
\hline
\end{tabular}
\end{table}

\begin{figure*}
\centering
\includegraphics[width=18cm, height=10cm]{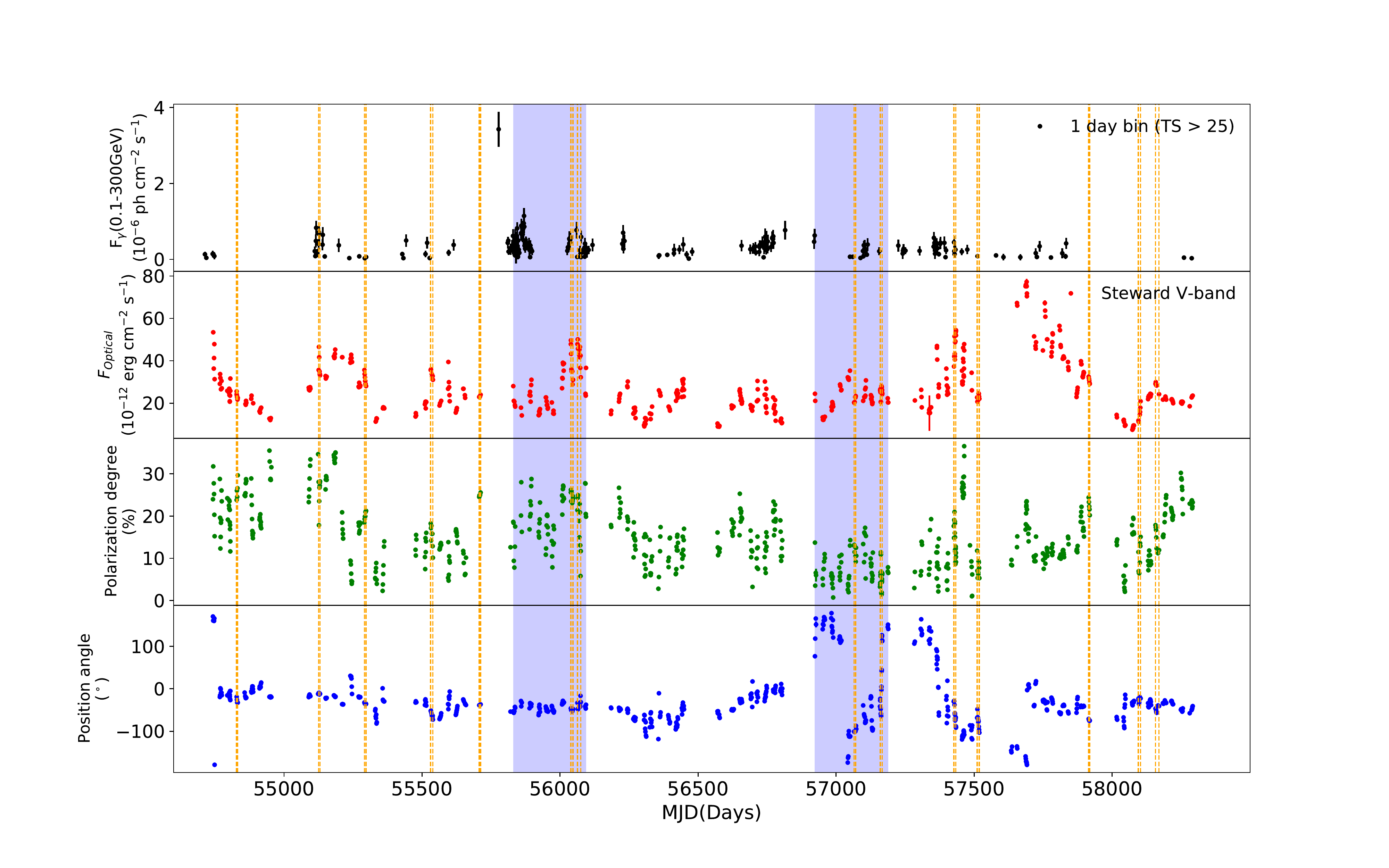}
\caption{\label{fig:multi_lc_oj287}Multi-wavelength light curve of OJ 287. The one-day binned $\gamma$-ray light curve is in the top panel. The optical V-band light curve is shown in the second panel. The third and fourth panels show the variations in PD and PA (corrected for the $180^{\circ}$ ambiguity), respectively.  The blue shaded regions reflect the cycles (on long-term timescales) where there was a positive or negative correlation between optical flux and PD. The epochs (on short-term timescales) where a correlation between optical flux and polarization was observed are indicated by dashed orange vertical lines.}
\end{figure*}

\begin{table*}
\centering
\caption{\label{tab:LTV_stats}Statistics of the observed flux and polarization properties on each observing cycles of all the sources. F$_{\gamma}$ is in units of 10$^{-6}$ ph cm$^{-2}$ s$^{-1}$. F$_{opt}$ is in units of 10$^{-11}$ erg cm$^{-2}$ s$^{-1}$. The polarization degree PD is in \%. Here, $R$ is the Spearman rank correlation coefficient, and $P$ is the probability of the null hypothesis.}
\resizebox{0.8\textwidth} {!}{  
\begin{tabular} {lccccccccccccccccc} \hline
Blazar  &  Cycle & Time-period  &  \multicolumn{3}{c}{F$_{\gamma}$} & \multicolumn{3}{c}{F$_{opt}$}     & \multicolumn{3}{c}{PD} & \multicolumn{3}{c}{F$_{opt}$ versus PD} & \multicolumn{3}{c} {F$_{\gamma}$ versus F$_{opt}$} \\
	&        & 		&       Min & Max & Average  &       Min & Max & Average        & Min  & Max  & Average  &         R & P   & Status & R &  P &  Status  \\ \hline
3C 66A 	& C2 & 55088$-$55383 &  0.02 &  0.62 & 0.25 $\pm$ 0.01 & 3.58 & 6.94 & 5.39 $\pm$ 0.01 & 2.15 & 19.68 & 9.53 $\pm$ 0.01  &	-0.28 &  0.05 & -  &  0.26 &  0.26 &  -\\ 
		& C3 & 55441$-$55769 &  0.03 &  0.56 & 0.19 $\pm$ 0.01 & 1.71 & 8.50 & 5.91 $\pm$ 0.02 & 8.11 & 19.59 & 15.40 $\pm$ 0.01  &	0.17 &  0.29 & -  &  0.18 &  0.54 &  -\\ 
		& C4 & 55800$-$56133 &  0.03 &  0.45 & 0.17 $\pm$ 0.02 & 1.73 & 3.58 & 2.35 $\pm$ 0.01 & 4.68 & 17.45 & 11.08 $\pm$ 0.01  &	-0.38 &  0.01 & -  &  0.00 &  0.00 &  -\\ 
		& C5 & 56184$-$56364 &  0.03 &  0.60 & 0.19 $\pm$ 0.02 & 2.62 & 4.59 & 3.58 $\pm$ 0.01 & 7.56 & 17.07 & 12.28 $\pm$ 0.01  &	0.01 &  0.93 & -  &  0.40 &  0.50 &  -\\ 
		& C6 & 56540$-$56863 &  0.01 & 0.38 & 0.15 $\pm$ 0.02 & 2.00 & 3.35 & 2.71 $\pm$ 0.01 & 2.47 & 13.85 & 6.74 $\pm$ 0.01  &	0.26 &  0.12 & -  &  0.00 &  0.00 &  -\\ 
		& C7 & 56920$-$57225 &  0.02 &  0.55 & 0.20 $\pm$ 0.02 & 2.43 & 4.46 & 2.97 $\pm$ 0.01 & 0.21 & 8.18 & 4.31 $\pm$ 0.01  &	-0.04 &  0.80 & -  &  0.00 &  0.00 &  -\\ 
		& C8 & 57282$-$57595 &  0.02 &  0.48 & 0.11 $\pm$ 0.01 & 2.10 & 2.76 & 2.48 $\pm$ 0.01 & 4.10 & 11.69 & 8.38 $\pm$ 0.02  &	-0.11 &  0.55 & -  &  0.00 &  0.00 &  -\\ 
		& C9 & 57630$-$57936 &  0.01 &  0.58 & 0.22 $\pm$ 0.02 & 2.26 & 3.58 & 2.87 $\pm$ 0.01 & 4.45 & 18.43 & 13.05 $\pm$ 0.01  &	-0.39 &  0.01 & -  &  0.61 &  0.15 &  -\\ 
		& C10 & 57996$-$58306 & 0.04 &  0.50 & 0.19 $\pm$ 0.03 & 1.53 & 2.66 & 2.22 $\pm$ 0.01 & 2.00 & 18.29 & 10.10 $\pm$ 0.01  &	-0.06 &  0.74 & -  &  0.00 &  0.00 &  -\\ 
          
AO 0235$+$164 	& C1 & 54743$-$54916 &  0.04 &  1.37 & 0.44 $\pm$ 0.01 & 0.12 & 1.32 & 0.51 $\pm$ 0.00 & 3.26 & 23.45 & 11.33 $\pm$ 0.06  &	0.42 &  0.00 & -  &  0.60 &  0.00 &  PC \\ 
		& C2 & 55089$-$55383 &  - &  - & - & 0.05 & 0.31 & 0.10 $\pm$ 0.00 & 0.58 & 11.95 & 4.38 $\pm$ 0.12  &	-0.24 &  0.16 & -  &  0.00 &  0.00 &  -\\ 
		& C3 & 55441$-$55769 &  - &  - & - & 0.03 & 0.07 & 0.05 $\pm$ 0.00 & 1.10 & 21.06 & 9.19 $\pm$ 0.09  &	-0.19 &  0.40 & -  &  0.00 &  0.00 &  -\\ 
		& C4 & 55800$-$56133 &  0.10 &  0.43 & 0.22 $\pm$ 0.06 & 0.05 & 0.10 & 0.07 $\pm$ 0.00 & 0.95 & 12.08 & 5.01 $\pm$ 0.09  &	0.71 &  0.01 & PC  &  0.00 &  0.00 &  -\\ 
		& C5 & 56214$-$56328 &  - &  - & - & 0.04 & 0.08 & 0.06 $\pm$ 0.00 & 2.04 & 25.30 & 7.10 $\pm$ 0.16  &	-0.10 &  0.87 & -  &  0.00 &  0.00 &  -\\ 
		& C6 & 56574$-$56658 &  0.09 &  0.35 & 0.22 $\pm$ 0.09 & 0.04 & 0.09 & 0.05 $\pm$ 0.00 & 2.39 & 6.67 & 4.56 $\pm$ 0.19  &	0.80 &  0.20 & -  &  0.00 &  0.00 &  -\\ 
		& C7 & 56920$-$57225 &  0.04 &  0.78 & 0.39 $\pm$ 0.02 & 0.07 & 0.58 & 0.17 $\pm$ 0.00 & 1.96 & 39.79 & 16.61 $\pm$ 0.05  &	0.16 &  0.36 & -  &  0.00 &  0.00 &  -\\ 
		& C8 & 57282$-$57598 &  0.04 &  0.91 & 0.34 $\pm$ 0.01 & 0.11 & 0.65 & 0.27 $\pm$ 0.00 & 1.05 & 32.11 & 13.53 $\pm$ 0.06  &	0.38 &  0.03 & -  &  0.50 &  0.03 &  PC\\ 
		& C9 & 57630$-$57826 &  0.03 &  0.50 & 0.22 $\pm$ 0.02 & 0.12 & 0.27 & 0.17 $\pm$ 0.00 & 2.91 & 27.21 & 9.17 $\pm$ 0.06  &	0.09 &  0.65 & -  &  0.00 &  0.00 &  -\\ 
		& C10 & 58016$-$58161 &  0.09 &  0.33 & 0.19 $\pm$ 0.07 & 0.07 & 0.17 & 0.12 $\pm$ 0.00 & 2.02 & 24.60 & 12.33 $\pm$ 0.08  &	-0.00 &  0.99 & -  &  0.00 &  0.00 &  -\\ 
   
S5 0716$+$714   & C1 & 54743$-$54952 &  0.03 &  0.61 & 0.20 $\pm$ 0.02 & 5.43 & 11.14 & 8.80 $\pm$ 0.02 & 3.30 & 15.51 & 8.78 $\pm$ 0.01  &	-0.43 &  0.06 & -  &  0.00 &  0.00 &  -\\ 
		& C2 & 55088$-$55336 &  0.04 &  0.71 & 0.27 $\pm$ 0.01 & 3.34 & 14.83 & 8.43 $\pm$ 0.01 & 0.54 & 27.96 & 12.00 $\pm$ 0.01  &	0.14 &  0.35 & -  &  0.08 &  0.77 &  -\\ 
		& C3 & 55563$-$55626 &  0.07 &  0.78 & 0.31 $\pm$ 0.02 & 3.66 & 13.40 & 7.14 $\pm$ 0.04 & 2.76 & 13.98 & 7.93 $\pm$ 0.01  &	0.24 &  0.44 & -  &  0.66 &  0.16 &  -\\ 
		& C4 & 55802$-$56014 &  0.03 &  1.24 & 0.34 $\pm$ 0.01 & 2.75 & 19.73 & 8.59 $\pm$ 0.02 & 0.95 & 21.40 & 10.17 $\pm$ 0.01  &	0.29 &  0.11 & -  &  0.42 &  0.05 &  -\\ 
		& C5 & 56184$-$56427 &  0.03 &  0.91 & 0.31 $\pm$ 0.01 & 3.76 & 18.16 & 6.88 $\pm$ 0.01 & 1.40 & 21.37 & 8.71 $\pm$ 0.01  &	0.06 &  0.72 & -  &  0.29 &  0.33 &  -\\ 
		& C6 & 56621$-$56804 &  0.04 &  1.38 & 0.36 $\pm$ 0.01 & 2.07 & 8.30 & 3.96 $\pm$ 0.01 & 1.89 & 17.59 & 9.85 $\pm$ 0.02  &	0.13 &  0.46 & -  &  0.00 &  0.00 &  -\\ 
		& C7 & 56922$-$57167 &  0.05 &  1.90 & 0.37 $\pm$ 0.01 & 4.82 & 30.41 & 16.36 $\pm$ 0.03 & 0.49 & 19.21 & 8.82 $\pm$ 0.01  &	-0.13 &  0.27 & -  &  0.73 &  0.00 &  PC\\ 
		& C8 & 57282$-$57311 &  0.04 &  0.32 & 0.16 $\pm$ 0.03 & 5.00 & 10.94 & 8.43 $\pm$ 0.06 & 2.10 & 19.77 & 13.45 $\pm$ 0.01  &	0.01 &  0.98 & -  &  0.00 &  0.00 &  -\\ 
		& C9 & 57630$-$57826 &  0.05 &  0.32 & 0.19 $\pm$ 0.03 & 3.79 & 10.26 & 6.26 $\pm$ 0.02 & 3.41 & 23.48 & 10.47 $\pm$ 0.02  &	0.73 &  0.03 & PC  &  0.00 &  0.00 &  -\\ 
		& C10 & 57996$-$58158 &  0.08 &  0.98 & 0.34 $\pm$ 0.01 & 4.28 & 10.45 & 8.65 $\pm$ 0.03 & 4.35 & 12.56 & 7.64 $\pm$ 0.02  &	-0.54 &  0.14 & -  &  0.52 &  0.29 &  -\\ 
		
OJ 287		& C1 & 54743$-$54952 &  0.08 &  0.14 & 0.11 $\pm$ 0.05 & 1.20 & 5.34 & 2.44 $\pm$ 0.00 & 11.64 & 35.49 & 22.24 $\pm$ 0.01  &	-0.19 &  0.19 & -  &  0.00 &  0.00 &  -\\ 
		& C2 & 55089$-$55363 &  0.03 &  0.83 & 0.33 $\pm$ 0.03 & 1.13 & 4.65 & 3.10 $\pm$ 0.00 & 2.29 & 35.05 & 19.73 $\pm$ 0.01  &	0.36 &  0.01 & -  &  0.83 &  0.04 &  PC\\ 
		& C3 & 55476$-$55712 &  0.03 &  0.43 & 0.23 $\pm$ 0.05 & 1.37 & 3.94 & 2.36 $\pm$ 0.00 & 4.72 & 25.57 & 13.62 $\pm$ 0.01  &	-0.00 &  0.99 & -  &  0.00 &  0.00 &  -\\ 
		& C4 & 55830$-$56094 &  0.06 &  1.14 & 0.43 $\pm$ 0.02 & 1.42 & 5.01 & 2.86 $\pm$ 0.00 & 5.81 & 28.78 & 19.67 $\pm$ 0.01  &	0.55 &  0.00 & PC  &  -0.61 &  0.04 &  NC\\ 
		& C5 & 56184$-$56449 &  0.09 &  0.70 & 0.29 $\pm$ 0.04 & 0.90 & 3.13 & 2.02 $\pm$ 0.00 & 2.80 & 26.69 & 13.32 $\pm$ 0.01  &	0.18 &  0.19 & -  &  0.00 &  0.00 &  -\\ 
		& C6 & 56570$-$56804 &  0.05 &  0.60 & 0.38 $\pm$ 0.03 & 0.89 & 3.05 & 1.86 $\pm$ 0.00 & 3.24 & 25.29 & 15.19 $\pm$ 0.01  &	0.29 &  0.04 & -  &  0.37 &  0.47 &  -\\ 
		& C7 & 56923$-$57189 &  0.04 &  0.39 & 0.20 $\pm$ 0.03 & 1.21 & 3.53 & 2.32 $\pm$ 0.00 & 0.73 & 17.23 & 7.41 $\pm$ 0.02  &	-0.53 &  0.00 & NC  &  0.00 &  0.00 &  -\\ 
		& C8 & 57285$-$57519 &  0.06 &  0.56 & 0.30 $\pm$ 0.02 & 1.53 & 5.44 & 3.22 $\pm$ 0.01 & 1.01 & 36.54 & 13.12 $\pm$ 0.01  &	0.23 &  0.02 & -  &  0.34 &  0.45 &  -\\ 
		& C9 & 57655$-$57919 &  0.05 &  0.42 & 0.17 $\pm$ 0.04 & 2.29 & 7.72 & 4.63 $\pm$ 0.01 & 7.52 & 24.40 & 14.68 $\pm$ 0.01  &	0.06 &  0.66 & -  &  0.00 &  0.00 &  -\\ 
		& C10 & 58016$-$58292 &  0.03 &  0.04 & 0.04 $\pm$ 0.01 & 0.74 & 2.99 & 1.86 $\pm$ 0.00 & 2.09 & 30.22 & 15.90 $\pm$ 0.01  &	0.25 &  0.06 & -  &  0.00 &  0.00 &  -\\ 
		           
MRK 421	& C1 & 54743$-$54952 &  0.01 &  0.49 & 0.14 $\pm$ 0.01 & 7.41 & 12.53 & 9.96 $\pm$ 0.03 & 0.63 & 6.12 & 3.41 $\pm$ 0.01  &	0.37 &  0.01 & -  &  0.12 &  0.59 &  -\\ 
		& C2 & 55125$-$55384 &  0.02 &  0.58 & 0.17 $\pm$ 0.01 & 8.35 & 15.48 & 11.12 $\pm$ 0.03 & 0.30 & 6.85 & 3.51 $\pm$ 0.00  &	-0.17 &  0.24 & -  &  0.34 &  0.06 &  -\\ 
		& C3 & 55477$-$55744 &  0.03 &  0.57 & 0.16 $\pm$ 0.01 & 10.32 & 21.57 & 15.47 $\pm$ 0.05 & 0.07 & 6.57 & 2.42 $\pm$ 0.00  &	0.44 &  0.00 & -  &  -0.12 &  0.63 &  -\\ 
		& C4 & 55858$-$56126 &  0.02 &  1.22 & 0.20 $\pm$ 0.01 & 9.77 & 23.43 & 15.72 $\pm$ 0.04 & 0.29 & 6.68 & 2.85 $\pm$ 0.00  &	0.18 &  0.18 & -  &  0.25 &  0.14 &  -\\ 
		& C5 & 56213$-$56449 &  0.03 &  0.81 & 0.25 $\pm$ 0.01 & 14.12 & 34.19 & 25.91 $\pm$ 0.06 & 0.64 & 11.44 & 3.52 $\pm$ 0.00  &	-0.38 &  0.00 & -  &  0.51 &  0.00 &  PC\\ 
		& C6 & 56571$-$56863 &  0.02 &  0.75 & 0.22 $\pm$ 0.01 & 9.24 & 24.09 & 15.41 $\pm$ 0.04 & 0.97 & 10.36 & 3.62 $\pm$ 0.00  &	0.34 &  0.00 & -  &  -0.00 &  0.97 &  -\\ 
		& C7 & 56924$-$57224 &  0.02 &  0.60 & 0.21 $\pm$ 0.01 & 10.71 & 29.50 & 17.54 $\pm$ 0.05 & 0.75 & 12.39 & 6.41 $\pm$ 0.01  &	0.22 &  0.07 & -  &  0.14 &  0.34 &  -\\ 
		& C8 & 57308$-$57593 &  0.02 &  0.67 & 0.17 $\pm$ 0.01 & 9.77 & 19.31 & 15.29 $\pm$ 0.05 & 0.39 & 4.82 & 2.21 $\pm$ 0.01  &	-0.50 &  0.00 & -  &  -0.03 &  0.87 &  -\\ 
		& C9 & 57687$-$57936 &  0.02 &  0.47 & 0.16 $\pm$ 0.01 & 7.08 & 10.14 & 8.18 $\pm$ 0.02 & 0.17 & 5.59 & 2.09 $\pm$ 0.01  &	-0.38 &  0.00 & -  &  -0.39 &  0.03 &  -\\ 
		& C10 & 58041$-$58306 &  0.02 &  0.84 & 0.22 $\pm$ 0.01 & 5.94 & 14.12 & 10.78 $\pm$ 0.03 & 0.05 & 4.28 & 1.86 $\pm$ 0.01  &	-0.14 &  0.17 & -  &  0.48 &  0.00 &  -\\ 
          
W Com 		& C1 & 54767$-$54953 &  0.011 &  0.44 & 0.20 $\pm$ 0.02 & 1.70 & 4.12 & 2.51 $\pm$ 0.02 & 3.29 & 17.36 & 9.49 $\pm$ 0.01  &	0.55 &  0.00 & PC  &  0.00 &  0.00 &  -\\ 
		& C2 & 55125$-$55384 &  0.12 & 0.50 & 0.31 $\pm$ 0.09 & 1.45 & 2.44 & 1.88 $\pm$ 0.01 & 7.09 & 20.67 & 13.13 $\pm$ 0.01  &	-0.45 &  0.00 & -  &  0.00 &  0.00 &  -\\ 
		& C3 & 55510$-$55769 &  0.04 &  0.17 & 0.10 $\pm$ 0.05 & 1.16 & 2.35 & 1.50 $\pm$ 0.01 & 1.91 & 14.57 & 7.52 $\pm$ 0.01  &	0.21 &  0.24 & -  &  0.00 &  0.00 &  -\\ 
		& C4 & 55891$-$56126 &  0.09 &  0.16 & 0.13 $\pm$ 0.04 & 0.74 & 2.06 & 1.48 $\pm$ 0.01 & 6.72 & 19.84 & 11.18 $\pm$ 0.01  &	0.18 &  0.29 & -  &  0.00 &  0.00 &  -\\ 
		& C5 & 56268$-$56448 &  0.09 &  0.16 & 0.13 $\pm$ 0.05 & 0.66 & 1.99 & 0.98 $\pm$ 0.01 & 7.57 & 22.30 & 13.43 $\pm$ 0.04  &	-0.28 &  0.15 & -  &  0.00 &  0.00 &  -\\ 
		& C6 & 56622$-$56860 &  - &  - & - & 0.80 & 1.36 & 0.95 $\pm$ 0.01 & 6.32 & 26.69 & 17.76 $\pm$ 0.01  &	0.07 &  0.60 & -  &  0.00 &  0.00 &  -\\ 
		& C7 & 56986$-$57219 &  - &  - & - & 0.88 & 1.67 & 1.22 $\pm$ 0.01 & 0.81 & 12.22 & 4.88 $\pm$ 0.02  &	-0.10 &  0.61 & -  &  0.00 &  0.00 &  -\\ 
		& C8 & 57365$-$57592 &  0.11 &  0.39 & 0.26 $\pm$ 0.05 & 0.74 & 1.30 & 1.01 $\pm$ 0.01 & 0.93 & 11.27 & 5.63 $\pm$ 0.02  &	-0.39 &  0.05 & -  &  0.00 &  0.00 &  -\\ 
		& C9 & 57691$-$57935 &  - &  - & - & 0.74 & 1.19 & 0.95 $\pm$ 0.01 & 3.31 & 16.27 & 9.85 $\pm$ 0.02  &	0.65 &  0.00 & PC  &  0.00 &  0.00 &  -\\ 
		& C10 & 58073$-$58306 &  - &  - & - & 1.24 & 3.01 & 1.98 $\pm$ 0.01 & 1.34 & 15.55 & 7.31 $\pm$ 0.01  &	0.27 &  0.08 & -  &  0.00 &  0.00 &  -\\ 
           
MRK 501	& C1 & 54745$-$54953 &  0.02 &  0.27 & 0.08 $\pm$ 0.02 & 5.13 & 6.34 & 5.56 $\pm$ 0.01 & 0.07 & 5.63 & 1.75 $\pm$ 0.01  &	0.64 &  0.00 & PC &  -0.90 &  0.01 &  NC\\ 
		& C2 & 55089$-$55384 &  0.02 &  0.09 & 0.04 $\pm$ 0.01 & 5.17 & 6.63 & 5.88 $\pm$ 0.01 & 0.81 & 4.54 & 2.51 $\pm$ 0.01  &	0.75 &  0.00 & PC  &  0.00 &  0.00 &  -\\ 
		& C3 & 55441$-$55769 &  0.02 &  0.50 & 0.08 $\pm$ 0.01 & 5.08 & 6.05 & 5.67 $\pm$ 0.01 & 1.65 & 5.14 & 3.21 $\pm$ 0.01  &	0.50 &  0.00 & -  &  -0.45 &  0.45 &  -\\ 
		& C4 & 55802$-$56133 &  0.02 &  0.47 & 0.12 $\pm$ 0.01 & 5.22 & 5.89 & 5.55 $\pm$ 0.01 & 1.68 & 5.91 & 3.45 $\pm$ 0.01  &	0.44 &  0.00 & -  &  -0.43 &  0.16 &  -\\ 
		& C5 & 56185$-$56480 &  0.01 &  0.43 & 0.13 $\pm$ 0.02 & 5.17 & 5.83 & 5.52 $\pm$ 0.01 & 1.15 & 3.93 & 2.31 $\pm$ 0.01  &	-0.19 &  0.23 & -  &  0.00 &  0.00 & -\\ 
		& C6 & 56540$-$56864 &  0.01 &  0.42 & 0.10 $\pm$ 0.01 & 5.42 & 6.39 & 5.88 $\pm$ 0.01 & 1.64 & 5.93 & 3.67 $\pm$ 0.01  &	0.57 &  0.00 & PC  &  -0.32 &  0.28 &  -\\ 
		& C7 & 56920$-$57224 &  0.01 &  0.51 & 0.11 $\pm$ 0.01 & 5.78 & 6.57 & 6.10 $\pm$ 0.01 & 1.00 & 5.48 & 2.46 $\pm$ 0.01  &	0.08 &  0.52 & -  &  0.45 &  0.12 &  -\\ 
		& C8 & 57282$-$57598 &  0.02 &  0.44 & 0.10 $\pm$ 0.01 & 5.89 & 6.82 & 6.24 $\pm$ 0.01 & 0.94 & 3.82 & 2.26 $\pm$ 0.01  &	0.17 &  0.25 & -  &  -0.06 &  0.86 &  -\\ 
		& C9 & 57630$-$57936 &  0.02 &  0.27 & 0.18 $\pm$ 0.04 & 5.22 & 6.28 & 5.74 $\pm$ 0.01 & 0.62 & 2.76 & 1.72 $\pm$ 0.01  &	-0.34 &  0.01 & -  &  0.00 &  0.00 &  -\\ 
		& C10 & 58014$-$58306 &  0.01 &  0.08 & 0.04 $\pm$ 0.02 & 5.32 & 5.89 & 5.63 $\pm$ 0.02 & 0.77 & 2.96 & 1.59 $\pm$ 0.01  &	0.20 &  0.25 & -  &  0.00 &  0.00 &  -\\           
1ES 1959$+$650	& C1 & 54745$-$54952  &  0.02 &  0.24 &  0.12 $\pm$ 0.05 & 3.04 & 4.43 & 3.54 $\pm$ 0.02 & 3.18 & 7.41 & 5.67 $\pm$ 0.04  &	0.08 &  0.75 & - &  0.00 &  0.00 &  -\\ 
		& C2 & 55089$-$55336  &  0.01 &  0.20 & 0.08 $\pm$ 0.02 & 1.95 & 3.79 & 2.83 $\pm$ 0.01 & 2.40 & 8.53 & 5.00 $\pm$ 0.01  &	0.83 &  0.00 & PC  &  0.00 &  0.00 &  - \\ 
		& C3 & 55707$-$55769  &  0.06 &  0.07 & 0.06 $\pm$ 0.03 & 2.98 & 3.33 & 3.14 $\pm$ 0.02 & 2.47 & 6.90 & 4.32 $\pm$ 0.02  &	0.36 &  0.43 & -  & 0.00 &  0.00 &  -\\ 
		& C4 & 55800$-$56133  &  0.02 &  0.30 &  0.10 $\pm$ 0.03 &  3.30 & 5.79 & 4.43 $\pm$ 0.02 & 2.49 & 6.51 & 4.66 $\pm$ 0.01  &	-0.82 &  0.00 & NC  &  0.00 & 0.00 & -\\ 
		& C5 & 56184$-$56477  &  0.03 &  0.12 &  0.07 $\pm$ 0.02 & 2.80 & 7.70 & 4.97 $\pm$ 0.03 & 2.05 & 3.91 & 3.01 $\pm$ 0.01  &	-0.02 &  0.96 & -  &  0.00 &  0.00 &  -\\ 
		& C6 & 56622$-$56863  &  0.03 &  0.26 &  0.11 $\pm$ 0.03 & 2.57 & 4.95 & 3.74 $\pm$ 0.02 & 2.73 & 6.00 & 3.93 $\pm$ 0.02  &	0.49 &  0.04 & -  &  0.00 &  0.00 &  -\\ 
		& C7 & 56920$-$57019 &  - &  - & - & 3.36 & 3.86 & 3.53 $\pm$ 0.02 & 2.08 & 5.10 & 3.82 $\pm$ 0.02  &	-0.54 &  0.11 & -  &  0.00 &  0.00 &  -\\ 
		& C8 & 57282$-$57598  &  0.01 &  0.49 &  0.14 $\pm$ 0.01 & 3.27 & 4.91 & 3.98 $\pm$ 0.02 & 0.37 & 4.66 & 2.46 $\pm$ 0.01  &	0.47 &  0.01 & -  &  -0.24 &  0.51 &  -\\ 
		& C9 & 57630$-$57826  &  0.02 &  0.35 &  0.11 $\pm$ 0.01 & 4.82 & 6.17 & 5.50 $\pm$ 0.03 & 1.14 & 5.00 & 2.71 $\pm$ 0.02  &	-0.64 &  0.06 & -  &  0.00 &  0.00 &  -\\ 
		& C10 & 57996$-$58293  &  0.01 &  0.38 &  0.12 $\pm$ 0.01 & 5.04 & 6.59 & 6.25 $\pm$ 0.06 & 2.15 & 5.08 & 3.86 $\pm$ 0.02  &	0.02 &  0.97 & -  &  0.00 &  0.00 &  -\\ 
		 
PKS 2155$-$304	& C1 & 54745$-$54804 &  0.04 &  0.43 &  0.23 $\pm$ 0.02 & 9.66 & 15.03 & 11.53 $\pm$ 0.02 & 1.90 & 10.09 & 6.33 $\pm$ 0.01  &	0.16 &  0.47 & -  &  0.59 &  0.03 &  PC\\ 
		& C2 & 55089$-$55383 &  0.01 &  0.63 &  0.14 $\pm$ 0.01 & 5.77 & 10.12 & 7.55 $\pm$ 0.02 & 0.92 & 8.11 & 4.27 $\pm$ 0.01  &	0.06 &  0.72 & -  &  0.32 &  0.29 &  -\\ 
		& C3 & 55441$-$55773 &  0.01 &  0.59 &  0.21 $\pm$ 0.01 & 5.26 & 22.96 & 12.13 $\pm$ 0.04 & 1.25 & 10.49 & 4.92 $\pm$ 0.01  &	0.21 &  0.24 & -  &  0.59 &  0.01 &  PC\\ 
		& C4 & 55800$-$56133 &  0.01 &  0.38 &  0.12 $\pm$ 0.01 & 4.22 & 9.23 & 6.42 $\pm$ 0.01 & 0.32 & 8.71 & 3.90 $\pm$ 0.01  &	-0.17 &  0.29 & -  &  0.12 &  0.78 &  -\\ 
		& C5 & 56184$-$56477 & 0.02 &  0.58 &  0.15 $\pm$ 0.01 & 5.66 & 10.31 & 7.57 $\pm$ 0.02 & 1.65 & 8.43 & 5.57 $\pm$ 0.01  &	0.17 &  0.39 & -  &  -0.31 &  0.45 &  -\\ 
		& C6 & 56539$-$56863 &  0.02 &  1.06 &  0.26 $\pm$ 0.01 & 4.67 & 12.16 & 6.74 $\pm$ 0.01 & 1.27 & 10.54 & 4.80 $\pm$ 0.01  &	0.61 &  0.00 & PC  &  0.7 &  0.19 &  -\\ 
		& C7 & 56920$-$57224 &  0.02 &  0.27 &  0.11 $\pm$ 0.01 & 3.00 & 9.31 & 4.83 $\pm$ 0.01 & 0.17 & 10.73 & 4.36 $\pm$ 0.01  &	0.80 &  0.00 & PC  &  0.00 &  0.00 &  -\\ 
		& C8 & 57282$-$57598 &  0.02 &  0.76 &  0.17 $\pm$ 0.01 & 6.87 & 12.05 & 8.85 $\pm$ 0.03 & 1.29 & 7.69 & 4.29 $\pm$ 0.01  &	0.17 &  0.46 & -  &  0.00 &  1.00 &  -\\ 
		& C9 & 57630$-$57936 &  0.01 &  0.61 &  0.19 $\pm$ 0.01 & 4.46 & 9.84 & 7.34 $\pm$ 0.02 & 2.11 & 19.06 & 10.03 $\pm$ 0.01  &	0.63 &  0.00 & PC  &  -0.22 &  0.53 &  -\\ 
		& C10 & 57996$-$58306 &  0.02 &  0.63 &  0.17 $\pm$ 0.01 & 5.77 & 12.05 & 7.77 $\pm$ 0.02 & 1.13 & 13.05 & 6.08 $\pm$ 0.01  &	0.77 &  0.00 & PC  &  0.58 &  0.05 &  PC\\ 
   
BL Lac          & C1 & 54743$-$54954 &  0.22 &  0.92 & 0.60 $\pm$ 0.06 & 3.68 & 6.22 & 4.69 $\pm$ 0.01 & 10.85 & 25.62 & 18.17 $\pm$ 0.02  &	-0.27 &  0.01 & -  &  0.00 &  0.00 &  -\\ 
		& C2 & 55089$-$55389 &  0.04 &  1.08 & 0.42 $\pm$ 0.02 & 3.20 & 10.13 & 5.83 $\pm$ 0.01 & 2.33 & 17.33 & 9.88 $\pm$ 0.01  &	-0.32 &  0.01 & -  &  -0.03 &  0.92 &  -\\ 
		& C3 & 55441$-$55774 &  0.14 &  1.42 & 0.61 $\pm$ 0.02 & 2.69 & 15.33 & 5.94 $\pm$ 0.01 & 2.14 & 23.18 & 12.24 $\pm$ 0.01  &	-0.74 &  0.00 & NC & 0.74 &  0.00 &  PC \\ 
		& C4 & 55800$-$56133 &  0.08 &  2.63 & 0.60 $\pm$ 0.01 & 5.41 & 17.44 & 8.96 $\pm$ 0.01 & 1.06 & 22.78 & 9.56 $\pm$ 0.01  &	0.38 &  0.00 & -  &  -0.11 &  0.47 &  -\\ 
		& C5 & 56179$-$56481 &  0.06 & 1.77 & 0.62 $\pm$ 0.03 & 1.55 & 8.35 & 3.76 $\pm$ 0.01 & 0.67 & 26.08 & 8.59 $\pm$ 0.01  &	0.32 &  0.00 & -  &  0.05 &  0.89 &  -\\ 
		& C6 & 56539$-$56864 &  0.03 &  1.23 & 0.50 $\pm$ 0.03 & 4.38 & 17.44 & 9.33 $\pm$ 0.01 & 0.71 & 22.36 & 7.32 $\pm$ 0.01  &	0.12 &  0.13 & -  &  0.47 &  0.01 &  -\\ 
		& C7 & 56920$-$57225 &  0.08 &  1.83 & 0.56 $\pm$ 0.02 & 4.67 & 18.94 & 9.99 $\pm$ 0.02 & 1.23 & 16.53 & 7.41 $\pm$ 0.01  &	-0.05 &  0.66 & -  &  0.76 &  0.00 &  PC\\ 
		& C8 & 57279$-$57598 &  0.04 &  1.67 & 0.48 $\pm$ 0.01 & 5.57 & 16.20 & 10.89 $\pm$ 0.02 & 0.88 & 13.07 & 5.49 $\pm$ 0.01  &	0.18 &  0.15 & -  &  0.17 &  0.29 &  -\\ 
		& C9 & 57630$-$57957 &  0.05 &  1.27 & 0.44 $\pm$ 0.01 & 4.07 & 15.33 & 10.67 $\pm$ 0.01 & 0.74 & 15.79 & 8.20 $\pm$ 0.01  &	0.44 &  0.00 & -  &  0.20 &  0.27 &  -\\ 
		& C10 & 58013$-$58312 &  0.05 &  1.79 & 0.53 $\pm$ 0.01 & 9.32 & 22.99 & 12.74 $\pm$ 0.02 & 0.62 & 16.81 & 7.31 $\pm$ 0.01  &	0.45 &  0.00 & -  &  0.35 &  0.01 &  -\\ 
		
1ES 2344$+$514  & C1 & 54743$-$54888 & - &  - & - & 1.57 & 2.17 & 2.00 $\pm$ 0.01 & 0.54 & 2.34 & 1.45 $\pm$ 0.01  &	-0.16 &  0.34 & -  &  0.00 & 0.00 &  -\\ 
		& C2 & 55090$-$55383 & - &  - & - & 1.65 & 2.23 & 2.04 $\pm$ 0.01 & 0.52 & 3.73 & 1.97 $\pm$ 0.01  &	0.44 &  0.01 & -  &  0.00 &  0.00 &  -\\ 
		& C3 & 55441$-$55740 & - &  - & - & 2.06 & 2.38 & 2.20 $\pm$ 0.02 & 1.13 & 4.54 & 3.17 $\pm$ 0.01  &	0.15 &  0.44 & -  &  0.00 &  0.00 &  -\\ 
		& C4 & 55800$-$56133 &  0.12 &  0.21 &  0.17 $\pm$ 0.07 & 1.98 & 2.52 & 2.26 $\pm$ 0.02 & 0.67 & 5.63 & 2.95 $\pm$ 0.02  &	0.40 &  0.06 & -  &  0.00 &  0.00 &  -\\ 
		& C5 & 56184$-$56421 & - &  - & - & 2.02 & 2.32 & 2.13 $\pm$ 0.02 & 0.90 & 3.81 & 2.17 $\pm$ 0.02  &	0.23 &  0.44 & -  &  0.00 &  0.00 &  -\\ 
		& C6 & 56547$-$56863 & - &  - & - &  2.00 & 2.32 & 2.17 $\pm$ 0.02 & 0.85 & 4.49 & 2.14 $\pm$ 0.02  &	0.19 &  0.57 & -  &  0.00 &  0.00 &  -\\ 
		& C7 & 56922$-$57188 &  0.02 &  0.06 &  0.05 $\pm$ 0.02 & 2.08 & 2.30 & 2.17 $\pm$ 0.02 & 0.45 & 3.81 & 2.46 $\pm$ 0.02  &	0.27 &  0.39 & -  &  0.00 &  0.00 &  -\\ 
		& C8 & 57306$-$57595 & - &  - & - & 2.13 & 2.59 & 2.33 $\pm$ 0.02 & 1.79 & 5.29 & 3.27 $\pm$ 0.02  &	0.20 &  0.43 & -  &  0.00 &  0.00 &  -\\ 
		& C9 & 57631$-$57784 & - &  - & - & 2.11 & 2.43 & 2.25 $\pm$ 0.02 & 1.33 & 5.20 & 3.49 $\pm$ 0.03  &	0.46 &  0.13 & -  &  0.00 &  0.00 &  -\\ 
		& C10 & 58014$-$58282 & - &  - & - & 1.98 & 2.32 & 2.17 $\pm$ 0.02 & 0.75 & 5.59 & 2.28 $\pm$ 0.03  &	0.32 &  0.24 & -  &  0.00 &  0.00 &  -\\   
\hline
\end{tabular}}
\end{table*}

\section{Reduction of multiwavelength data}\label{sec:data}
To look for the correlation between optical V-band flux and polarization variations we used the publicly available data from the Steward observatory\footnote{http://james.as.arizona.edu/$\sim$psmith/Fermi} spanning a period of $\sim$ 10 years from 2008-2018. And to look for the correlation between optical and $\gamma$-ray flux variations on a long-term timescale, we used $\gamma$-ray data from $\it{Fermi}$, that spans the same period as the optical data.

\subsection{Optical data reduction}
 We obtained photometric data in the optical V-band and R-band, as well as polarimetric data in the optical V-band, from Steward Observatory for 11 BL Lac sources listed in Table \ref{tab:src_list}. These observations were made with SPOL, which is a moderate resolution (R $\sim$ 300$-$1000) spectropolarimeter. A $\lambda$/4 plate is used to measure circular polarization and a $\lambda$/2 plate is used to measure linear polarization in the telescope. This instrument also provides the imaging photometric observations over a narrow field of view using the 2.3 m Bok Telescope on Kitt Peak and 1.5 m Kuiper Telescope on Mt. Bigelow.  The observation and reduction procedures of this publicly archival data are given in \cite{2009arXiv0912.3621S}. To convert the magnitude into flux we followed the same procedure as described in \cite{2022MNRAS.510.1809P}. Firstly, the magnitudes were corrected for galactic extinction ($A_{\lambda}$), the values of which were taken from NED \footnote{https://ned.ipac.caltech.edu/}. We then converted the extinction corrected magnitudes into their flux values using the coefficient given in \cite{1998A&A...333..231B}. We note that we have not made any corrections for the host galaxy light.

\subsection{$\gamma$-ray data reduction}
We collected $\gamma$-ray data in the energy range 0.1-300 GeV for 11 BL Lac sources detected by the $\it{Fermi}$ Large Area Telescope ($\it{Fermi}$-LAT; \citealt{2009ApJ...697.1071A}). To generate the one-day binned $\gamma$-ray light curves of these sources we followed the same strategy given in \cite{2021MNRAS.504.1772R}. If the test statistics (TS) $\geq$ 25 (which corresponds to 5$\sigma$; \citealt{1996ApJ...461..396M}), the source is considered to be detected in the one-day binned $\gamma$-ray light curve.
\begin{table*}
\centering
\caption{\label{tab:STV_stats} Statistical analysis of observed flux and polarization parameters on short-term timescales. Here '-' denotes that the source was not detected by Fermi at the epoch. The rest of the information is the same as it is in the caption of Table \ref{tab:LTV_stats}.}
\resizebox{1.0\textwidth} {!}{  
\begin{tabular} {lcccccccccccccc} \hline\hline
Blazar 	& Epoch & Time-period    & \multicolumn{3}{c}{F$_{\gamma}$ ($10^{-6}$ ph $cm^{2}$ $s^{-1}$)}  &  \multicolumn{3}{c}{F$_{opt}$ ($10^{-12}$ erg $cm^{2}$ $s^{-1}$)}     & \multicolumn{3}{c}{PD (\%)} & \multicolumn{3}{c}{PA (degree)} \\
       	&       &                & Min & Max & Average               & Min  & Max  & Average              & Min  & Max  & Average  & Min & Max  & Average        \\ \hline
3C 66A  	& A & 55240$-$55246 & 0.07 & 0.17 & 0.12 $\pm$ 0.06 & 54.10 & 68.20 & 60.04 $\pm$ 0.40 & 2.15 & 10.66 & 7.01 $\pm$ 0.01 & 25.20 & 38.60 & 31.51 $\pm$ 0.08 \\ 
		& B & 55531$-$55539 & 0.11 & 0.31 & 0.21 $\pm$ 0.05 & 53.10 & 72.00 & 60.60 $\pm$ 0.53 & 11.27 & 18.49 & 15.37 $\pm$ 0.01 & 31.10 & 36.50 & 34.21 $\pm$ 0.03 \\ 
        & C & 55922$-$55927 & - & - & - & 24.10 & 27.10 & 25.40 $\pm$ 0.19 & 6.81 & 10.93 & 8.10 $\pm$ 0.02 & 10.80 & 14.10 & 12.95 $\pm$ 0.08 \\ 
        & D & 56920$-$56927 & - & - & - & 24.30 & 27.10 & 25.41 $\pm$ 0.22 & 3.87 & 7.22 & 5.17 $\pm$ 0.02 & -25.20 & -10.10 & -16.68 $\pm$ 0.11 \\ 
		 
AO 0235$+$164   & A & 54743$-$54748 & 0.40 & 0.97 & 0.71 $\pm$ 0.06 & 4.06 & 9.13 & 6.80 $\pm$ 0.00 & 13.61 & 23.45 & 17.53 $\pm$ 0.03 & 80.80 & 118.60 & 98.24 $\pm$ 0.06 \\ 
 
S5 0716$+$714   & A & 55594$-$55600 & 0.20 & 0.20 & 0.20 $\pm$ 0.08 & 36.60 & 44.80 & 41.07 $\pm$ 0.15 & 3.18 & 10.38 & 5.85 $\pm$ 0.01 & -160.10 & -33.40 & -92.10 $\pm$ 0.08 \\ 
                & B & 56690$-$56697 & - & - & - & 38.30 & 59.60 & 46.45 $\pm$ 0.18 & 1.89 & 17.28 & 9.62 $\pm$ 0.02 & -163.80 & -86.50 & -111.82 $\pm$ 0.14 \\   
OJ 287	& A & 54828$-$54832 & - & - & - & 21.70 & 25.60 & 23.37 $\pm$ 0.09 & 23.81 & 29.65 & 26.13 $\pm$ 0.02 & -32.60 & -19.00 & -28.48 $\pm$ 0.04 \\ 
        & B & 55125$-$55130 & 0.40 & 0.71 & 0.59 $\pm$ 0.09 & 33.10 & 46.50 & 37.60 $\pm$ 0.14 & 17.84 & 28.22 & 25.63 $\pm$ 0.02 & -13.40 & -9.70 & -11.30 $\pm$ 0.04 \\ 
        & C & 55291$-$55297 & 0.03 & 0.06 & 0.04 $\pm$ 0.02 & 28.00 & 35.60 & 31.06 $\pm$ 0.11 & 18.47 & 21.24 & 20.13 $\pm$ 0.02 & -36.20 & -32.20 & -34.76 $\pm$ 0.04 \\ 
        & D & 55531$-$55539 & - & - & - & 31.00 & 36.00 & 33.60 $\pm$ 0.15 & 10.14 & 18.25 & 15.31 $\pm$ 0.02 & -72.60 & -51.50 & -61.00 $\pm$ 0.04 \\ 
        & E & 55707$-$55712 & - & - & - & 22.90 & 24.00 & 23.24 $\pm$ 0.10 & 24.52 & 25.57 & 24.97 $\pm$ 0.02 & -38.70 & -36.80 & -38.06 $\pm$ 0.04 \\  
        & F & 56039$-$56047 & - & - & - & 28.80 & 49.60 & 37.50 $\pm$ 0.12 & 23.08 & 26.22 & 24.81 $\pm$ 0.02 & -53.40 & -46.00 & -48.14 $\pm$ 0.03 \\ 
        & G & 56063$-$56075 & 0.06 & 0.24 & 0.15 $\pm$ 0.05 & 32.20 & 50.10 & 43.12 $\pm$ 0.17 & 5.81 & 24.94 & 17.88 $\pm$ 0.02 & -48.30 & -16.70 & -36.79 $\pm$ 0.05 \\ 
        & H & 57065$-$57071 & - & - & - & 20.10 & 23.30 & 21.63 $\pm$ 0.10 & 9.21 & 13.20 & 11.68 $\pm$ 0.03 & -102.00 & -89.30 & -98.77 $\pm$ 0.07 \\ 
        & I & 57160$-$57167 & - & - & - & 20.50 & 28.00 & 25.39 $\pm$ 0.06 & 1.47 & 11.46 & 5.43 $\pm$ 0.01 & -64.30 & 126.00 & 9.62 $\pm$ 0.11 \\ 
        & J & 57427$-$57434 & 0.14 & 0.45 & 0.26 $\pm$ 0.07 & 37.30 & 54.40 & 47.36 $\pm$ 0.15 & 8.65 & 20.97 & 13.37 $\pm$ 0.01 & -91.60 & -28.60 & -65.39 $\pm$ 0.03 \\ 
        & K & 57511$-$57519 & 0.08 & 0.08 & 0.08 $\pm$ 0.04 & 20.70 & 24.70 & 22.59 $\pm$ 0.10 & 5.32 & 11.80 & 7.53 $\pm$ 0.03 & -103.30 & -48.20 & -76.99 $\pm$ 0.15 \\ 
        & L & 57915$-$57919 & - & - & - & 29.10 & 32.50 & 30.70 $\pm$ 0.21 & 20.20 & 24.40 & 22.36 $\pm$ 0.04 & -75.40 & -70.80 & -73.80 $\pm$ 0.04 \\ 
        & M & 58095$-$58103 & - & - & - & 11.20 & 20.90 & 15.37 $\pm$ 0.07 & 6.47 & 15.09 & 10.25 $\pm$ 0.05 & -35.70 & -19.40 & -25.65 $\pm$ 0.22 \\ 
        & N & 58157$-$58170 & - & - & - & 24.20 & 29.90 & 28.02 $\pm$ 0.12 & 11.21 & 17.86 & 14.62 $\pm$ 0.02 & -56.60 & -39.40 & -47.83 $\pm$ 0.06 \\ 
MRK 421	& A & 54743$-$54748 & 0.25 & 0.25 & 0.25 $\pm$ 0.13 & 111.00 & 123.00 & 114.60 $\pm$ 0.95 & 2.80 & 5.17 & 3.74 $\pm$ 0.01 & 59.60 & 120.20 & 107.06 $\pm$ 0.13 \\ 
        & B & 54911$-$54917 & 0.08 & 0.16 & 0.13 $\pm$ 0.04 & 96.80 & 112.00 & 103.97 $\pm$ 1.06 & 2.43 & 4.55 & 3.98 $\pm$ 0.02 & 126.60 & 132.20 & 128.65 $\pm$ 0.15 \\ 
        & C & 54947$-$54952 & 0.17 & 0.27 & 0.21 $\pm$ 0.05 & 100.00 & 112.00 & 106.00 $\pm$ 0.88 & 2.67 & 4.94 & 3.93 $\pm$ 0.01 & 139.20 & 143.80 & 140.82 $\pm$ 0.12 \\  
        & D & 56772$-$56781 & 0.19 & 0.62 & 0.31 $\pm$ 0.03 & 145.00 & 188.00 & 158.46 $\pm$ 0.59 & 1.80 & 5.19 & 3.09 $\pm$ 0.01 & 21.60 & 51.50 & 38.02 $\pm$ 0.07 \\ 
        & E & 57011$-$57019 & 0.08 & 0.45 & 0.20 $\pm$ 0.03 & 171.00 & 214.00 & 194.00 $\pm$ 2.08 & 6.32 & 12.38 & 10.18 $\pm$ 0.01 & -29.90 & -20.40 & -25.39 $\pm$ 0.04 \\ 
        & F & 57098$-$57108 & 0.02 & 0.33 & 0.18 $\pm$ 0.03 & 129.00 & 153.00 & 144.50 $\pm$ 1.09 & 3.55 & 7.02 & 4.86 $\pm$ 0.02 & -18.50 & 5.60 & -6.36 $\pm$ 0.10 \\ 
        & G & 57187$-$57195 & 0.12 & 0.57 & 0.34 $\pm$ 0.04 & 168.00 & 226.00 & 204.62 $\pm$ 1.34 & 0.75 & 5.82 & 3.38 $\pm$ 0.01 & -108.00 & -9.30 & -39.99 $\pm$ 0.26 \\ 
        & H & 57456$-$57464 & 0.11 & 0.28 & 0.18 $\pm$ 0.05 & 168.00 & 193.00 & 185.11 $\pm$ 1.32 & 0.43 & 2.72 & 1.40 $\pm$ 0.01 & -173.00 & -110.20 & -139.01 $\pm$ 0.47 \\ 
        & I & 57887$-$57898 & 0.03 & 0.27 & 0.09 $\pm$ 0.02 & 76.90 & 88.30 & 82.02 $\pm$ 0.64 & 0.26 & 3.55 & 2.02 $\pm$ 0.02 & -179.80 & 177.50 & 18.08 $\pm$ 0.79 \\ 
        & J & 58130$-$58141 & 0.09 & 0.48 & 0.30 $\pm$ 0.05 & 108.00 & 130.00 & 119.87 $\pm$ 0.64 & 0.20 & 2.15 & 1.21 $\pm$ 0.01 & 0.30 & 106.30 & 39.33 $\pm$ 0.36 \\ 
        & K & 58249$-$58256 & 0.18 & 0.44 & 0.33 $\pm$ 0.05 & 93.30 & 126.00 & 109.88 $\pm$ 0.72 & 2.11 & 4.04 & 3.27 $\pm$ 0.02 & 5.40 & 27.80 & 16.73 $\pm$ 0.15 \\ 
        & L & 58281$-$58292 & 0.03 & 0.22 & 0.13 $\pm$ 0.03 & 59.40 & 76.90 & 64.18 $\pm$ 0.48 & 1.94 & 3.58 & 2.71 $\pm$ 0.03 & -5.20 & 43.50 & 26.68 $\pm$ 0.35 \\ 
W Com 	& A & 56741$-$56749 & - & - & - & 7.99 & 10.30 & 8.78 $\pm$ 0.11 & 18.52 & 23.19 & 21.49 $\pm$ 0.03 & 68.20 & 79.00 & 72.53 $\pm$ 0.04 \\ 
        & B & 57887$-$57897 & - & - & - & 9.88 & 11.90 & 11.10 $\pm$ 0.15 & 11.96 & 16.23 & 14.34 $\pm$ 0.04 & 61.70 & 78.00 & 68.41 $\pm$ 0.09 \\    
MRK 501 & A & 56951$-$56959 & 0.11 & 0.30 & 0.19 $\pm$ 0.05 & 61.10 & 65.70 & 63.67 $\pm$ 0.36 & 1.84 & 3.63 & 2.72 $\pm$ 0.02 & -81.20 & -53.20 & -69.64 $\pm$ 0.22 \\ 
1ES 1959$+$650  & A & 55270$-$55275 & - & - & - & 19.50 & 23.70 & 22.52 $\pm$ 0.21 & 3.29 & 5.20 & 3.93 $\pm$ 0.03 & 140.20 & 153.70 & 148.44 $\pm$ 0.22 \\ 
                & B & 55291$-$55297 & - & - & - & 20.60 & 23.10 & 21.21 $\pm$ 0.15 & 3.10 & 4.51 & 3.59 $\pm$ 0.03 & 144.80 & 150.70 & 148.06 $\pm$ 0.21 \\ 
                & C & 57306$-$57311 & 0.10 & 0.19 & 0.15 $\pm$ 0.04 & 36.90 & 40.40 & 38.59 $\pm$ 0.27 & 0.37 & 3.13 & 1.79 $\pm$ 0.02 & 118.90 & 163.40 & 136.36 $\pm$ 0.49 \\ 
PKS 2155$-$304  & A & 55123$-$55130 & 0.10 & 0.16 & 0.13 $\pm$ 0.05 & 57.70 & 66.20 & 63.36 $\pm$ 0.21 & 1.26 & 5.98 & 3.81 $\pm$ 0.01 & 26.90 & 67.30 & 52.05 $\pm$ 0.16 \\ 
                & B & 55357$-$55363 & 0.17 & 0.17 & 0.17 $\pm$ 0.11 & 84.90 & 101.00 & 93.13 $\pm$ 0.90 & 3.03 & 8.11 & 5.23 $\pm$ 0.01 & 67.40 & 99.80 & 90.29 $\pm$ 0.08 \\ 
                & C & 55510$-$55515 & 0.17 & 0.37 & 0.26 $\pm$ 0.06 & 155.00 & 170.00 & 160.83 $\pm$ 0.60 & 2.99 & 6.38 & 4.29 $\pm$ 0.01 & 60.40 & 101.40 & 72.33 $\pm$ 0.08 \\ 
                & D & 57716$-$57724 & 0.22 & 0.27 & 0.24 $\pm$ 0.08 & 44.60 & 57.10 & 50.97 $\pm$ 0.24 & 2.11 & 12.51 & 6.80 $\pm$ 0.03 & -115.40 & -86.10 & -99.87 $\pm$ 0.18 \\ 
                & E & 58041$-$58048 & 0.10 & 0.29 & 0.18 $\pm$ 0.03 & 70.60 & 76.70 & 73.07 $\pm$ 0.33 & 3.37 & 9.12 & 6.21 $\pm$ 0.02 & -127.70 & -114.00 & -122.63 $\pm$ 0.09 \\ 
                & F & 58250$-$58256 & 0.14 & 0.14 & 0.14 $\pm$ 0.06 & 62.70 & 74.60 & 67.28 $\pm$ 0.48 & 3.28 & 4.57 & 3.97 $\pm$ 0.03 & -142.90 & -110.50 & -128.73 $\pm$ 0.21 \\ 
BL Lac	& A & 54743$-$54748 & - & - & - & 41.80 & 48.50 & 45.98 $\pm$ 0.14 & 14.36 & 19.29 & 17.45 $\pm$ 0.11 & 17.30 & 24.30 & 19.97 $\pm$ 0.22 \\ 
        & B & 54767$-$54774 & - & - & - & 48.90 & 62.10 & 55.54 $\pm$ 0.14 & 12.29 & 20.11 & 16.71 $\pm$ 0.01 & 13.90 & 29.10 & 22.14 $\pm$ 0.03 \\ 
        & C & 55383$-$55389 & - & - & - & 45.00 & 65.10 & 55.15 $\pm$ 0.21 & 13.28 & 17.15 & 15.30 $\pm$ 0.03 & 0.40 & 12.70 & 8.42 $\pm$ 0.05 \\ 
        & D & 56125$-$56133 & 0.18 & 1.74 & 0.74 $\pm$ 0.06 & 84.20 & 174.00 & 126.26 $\pm$ 0.38 & 10.42 & 22.78 & 15.36 $\pm$ 0.01 & 5.80 & 33.60 & 19.24 $\pm$ 0.03 \\ 
        & E & 56325$-$56332 & - & - & - & 15.50 & 23.20 & 19.13 $\pm$ 0.22 & 2.14 & 9.73 & 5.71 $\pm$ 0.05 & -179.80 & -136.00 & -162.08 $\pm$ 0.39 \\ 
        & F & 56772$-$56781 & 0.40 & 0.40 & 0.40 $\pm$ 0.21 & 52.20 & 111.00 & 81.31 $\pm$ 0.32 & 6.14 & 12.46 & 9.12 $\pm$ 0.02 & 33.00 & 67.20 & 50.89 $\pm$ 0.07 \\ 
        & G & 57098$-$57108 & 0.81 & 1.33 & 1.07 $\pm$ 0.25 & 129.00 & 176.00 & 158.50 $\pm$ 0.67 & 1.23 & 7.20 & 4.55 $\pm$ 0.02 & 29.20 & 92.90 & 62.71 $\pm$ 0.22 \\ 
        & H & 57125$-$57133 & 0.50 & 0.86 & 0.61 $\pm$ 0.08 & 108.00 & 145.00 & 128.44 $\pm$ 0.60 & 2.51 & 15.03 & 6.21 $\pm$ 0.02 & -14.10 & 48.70 & 11.94 $\pm$ 0.14 \\ 
        & I & 57279$-$57285 & 0.39 & 0.64 & 0.48 $\pm$ 0.07 & 89.80 & 140.00 & 111.05 $\pm$ 0.93 & 3.33 & 13.07 & 8.37 $\pm$ 0.03 & -35.40 & -13.60 & -26.87 $\pm$ 0.12 \\ 
        & J & 57630$-$57637 & 0.31 & 0.31 & 0.31 $\pm$ 0.14 & 40.70 & 56.20 & 46.53 $\pm$ 0.16 & 0.74 & 9.30 & 4.57 $\pm$ 0.03 & -126.50 & -47.30 & -92.44 $\pm$ 0.53 \\ 
        & K & 57749$-$57766 & 0.11 & 0.45 & 0.27 $\pm$ 0.05 & 93.20 & 117.00 & 103.74 $\pm$ 0.36 & 5.52 & 12.61 & 8.51 $\pm$ 0.02 & -13.50 & 13.40 & -1.17 $\pm$ 0.08 \\ 
        & L & 57886$-$57898 & 0.12 & 0.45 & 0.29 $\pm$ 0.05 & 89.00 & 123.00 & 102.98 $\pm$ 0.54 & 5.18 & 11.66 & 7.99 $\pm$ 0.04 & -45.30 & 3.10 & -14.64 $\pm$ 0.20 \\ 
        & M & 58249$-$58256 & 0.29 & 0.84 & 0.57 $\pm$ 0.09 & 106.00 & 136.00 & 118.38 $\pm$ 0.39 & 4.20 & 9.14 & 6.54 $\pm$ 0.02 & -13.10 & 20.30 & -0.80 $\pm$ 0.09 \\ 
 
\hline
\end{tabular}}
\end{table*}

\begin{figure*}
\centering
\includegraphics[width=18cm, height=15cm]{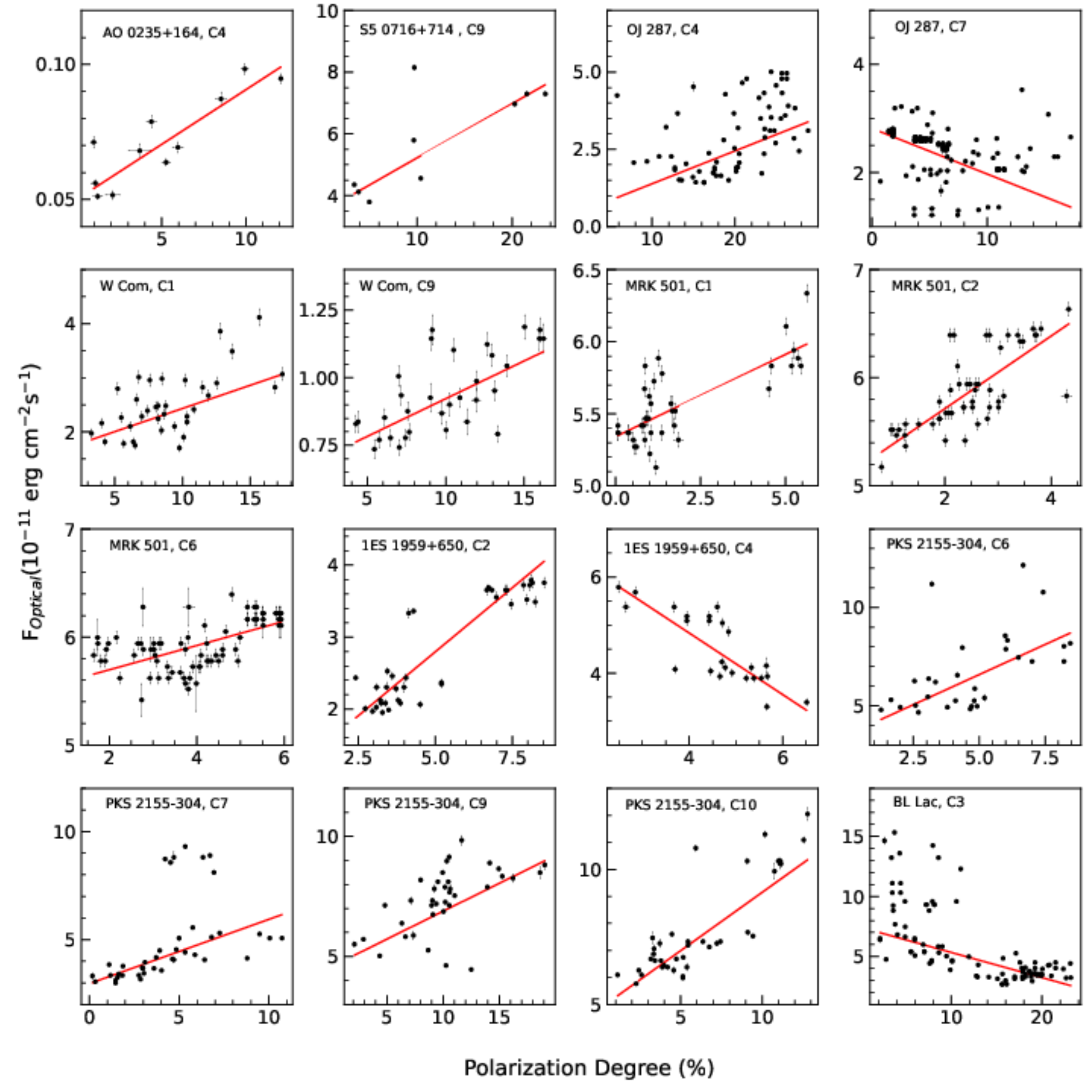}
\caption{Degree of polarization vs optical V-band flux on long-term timescales. Each panel includes the name of the source and the observing cycle. The solid red line is the linear least squares fit to the data.}
\label{fig:pol_flx_cor1}
\end{figure*}

\begin{figure*}
\centering
\includegraphics[width=18cm, height=12cm]{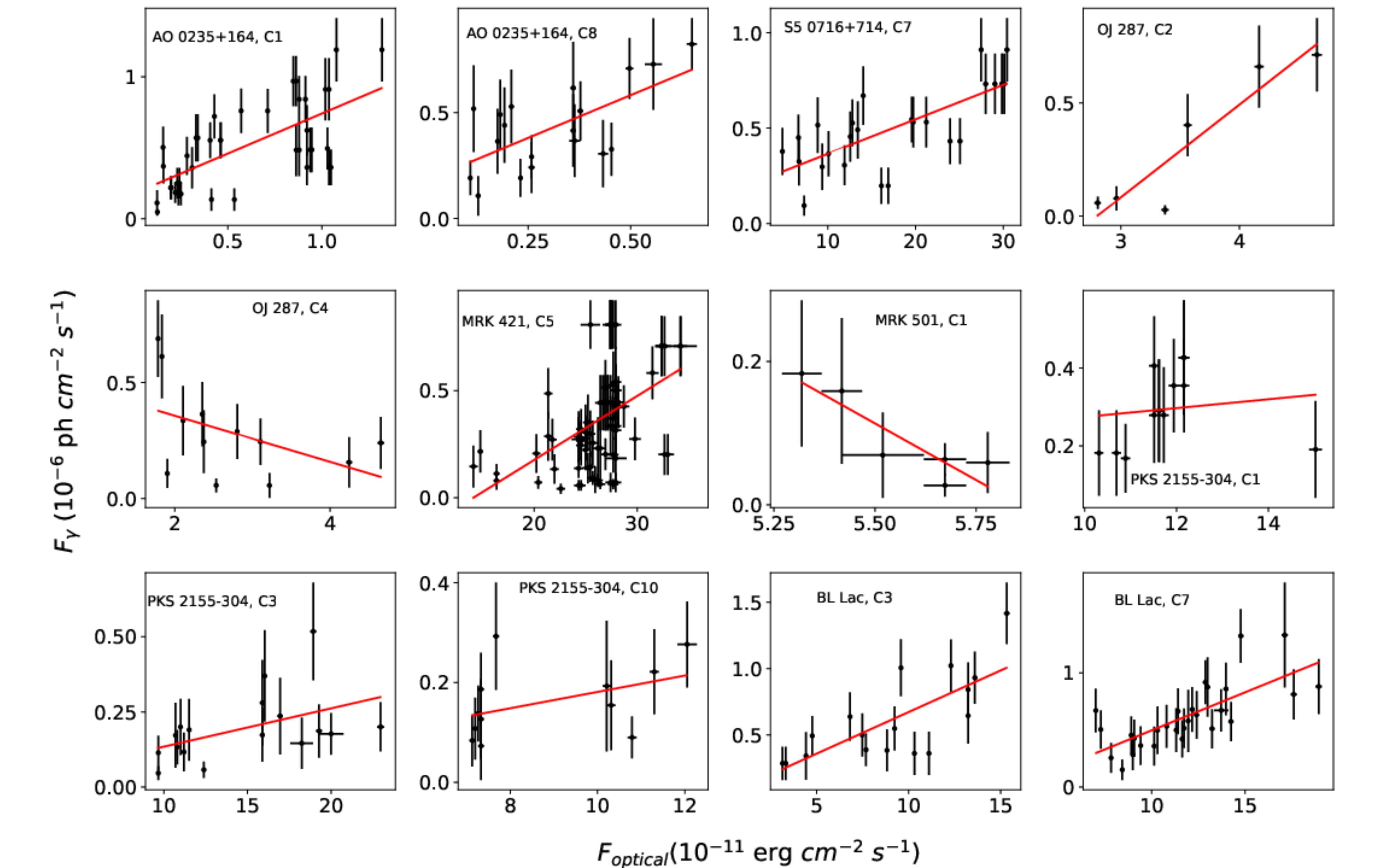}
\caption{ $\gamma$-ray flux vs optical V-band flux on long-term timescales. Each panel includes the name of the source and the observing cycle. The solid red line is the linear least squares fit to the data.}
\label{fig:opt_gamma_cor1}
\end{figure*}

\section{Multiwavelength Lightcurves} \label{sec:lightcurve}
 The multiwavelength light curve for the source OJ 287 is shown in Figure \ref{fig:multi_lc_oj287}, that include $\gamma$-ray, optical V-band, PD and electric vector position angle (PA) from 2008$-$2018 (MJD 54743$-$58306). For the other sources, the multiwavelength light curves are available in the electronic version. To resolve the $180^{\circ}$ ambiguity of the PA measurement we used the same approach as given in \cite{2022MNRAS.510.1809P}, and also described in \cite{2010Natur.463..919A} and \cite{2015MNRAS.453.1669B}. According to this the timing variation in the PA should be gradual and smooth, resulting in minimal PA variations between consecutive measurements. i.e. $-90^{\circ}$ $\leq$ $\Delta\theta$ $\leq$ $90^{\circ}$. Here $\Delta\theta$ is PA variation and defined as $|\theta_{n}-\theta_{n-1}|$-$\sqrt{\sigma(\theta_{n})^2+\sigma(\theta_{n-1})^2}$, where $\theta_{n-1}$ and $\theta_{n}$ are the $(n-1)^{th}$ and n$^{th}$ values of PA and $\sigma(\theta_{n-1})$, $\sigma(\theta_{n})$ are the corresponding errors of the PA values. If $\Delta\theta$ $>$ $90^{\circ}$, we shifted $\theta_{n}$ to $\theta_{n}$$-$$180^{\circ}$ and for $\Delta\theta$ $<$ $-90^{\circ}$, we shifted $\theta_{n}$ to $\theta_{n}$+$180^{\circ}$. And if $-90^{\circ}$ $\leq$ $\Delta\theta$ $\leq$ $90^{\circ}$, we left $\theta_{n}$ unchanged.

\begin{table}
\centering
\caption{\label{tab:epoch_cor} Correlation analysis results for optical flux versus PD on short-term timescales. Here, PC denotes a positive correlation between optical flux and PD, whereas NC denotes an negative correlation between optical flux and PD.}
\resizebox{0.45\textwidth} {!}{ 
\begin{tabular} {lccccccccccccc} \hline
Blazar		&  Epoch  & \multicolumn{3}{c}{F$_{opt}$ versus PD} \\
		&                                          & R        & P     & Remarks  \\ \hline  
3C 66A	& A   & -0.86 & 0.01 & NC \\ 
        & B   & 0.84 & 0.02  & PC \\ 
        & C   & 0.90 & 0.01  & PC \\ 
        & D   & 0.72 & 0.04  & PC \\ 

AO 0235$+$164  & A   & 0.76 & 0.00 & PC \\ 
  
S5 0716$+$714   & A  & 0.89 & 0.02 & PC \\ 
                & B  & 0.97 & 0.00 & PC \\ 

OJ 287	& A  & -0.87 & 0.02 & NC \\ 
        & B  & -0.77 & 0.04 & NC \\ 
        & C  & -1.00 & 0.00 & NC \\ 
        & D  & 0.89 & 0.02 & PC \\ 
        & E  & 0.97 & 0.00 & PC \\ 
        & F  & 0.65 & 0.02 & PC \\ 
        & G  & 0.65 & 0.04 & PC \\ 
        & H  & -0.83 & 0.04 & NC \\ 
        & I  & -0.93 & 0.00 & NC \\ 
        & J  & -0.67 & 0.00 & NC \\ 
        & K  & 0.72 & 0.04 & PC \\ 
        & L  & 1.00 & 0.00 & PC \\ 
        & M  & 0.89 & 0.02 & PC \\ 
        & N  & 0.90 & 0.04 & PC \\    
		     
MRK 421	& A  & -0.89 & 0.04 & NC \\ 
        & B  & 0.82 & 0.04 & PC \\ 
        & C  & -1.00 & 0.00 & NC \\ 
        & D  & 0.76 & 0.00 & PC \\ 
        & E  & 0.90 & 0.04 & PC \\ 
        & F  & -0.83 & 0.04 & NC \\ 
        & G  & 0.81 & 0.02 & PC \\ 
        & H  & -0.88 & 0.00 & NC \\ 
        & I  & -0.74 & 0.02 & NC \\ 
        & J  & 0.65 & 0.00 & PC \\ 
        & K  & 0.77 & 0.03 & PC \\ 
        & L  & -0.88 & 0.02 & NC \\        
		 
W Com	& A  & -0.84 & 0.00 & NC \\ 
        & B  & 0.72 & 0.04 & PC \\         
		  
MRK 501	 & A  & 0.83 & 0.04 & PC \\

1ES 1959$+$650  & A  & 0.90 & 0.04 & PC \\ 
                & B  & -0.77 & 0.04 & NC \\ 
                & C  & 0.63 & 0.00 & PC \\ 

PKS 2155$-$304  & A  & 0.75 & 0.03 & PC \\ 
                & B  & 0.85 & 0.01 & PC \\ 
                & C  & 0.81 & 0.05 & PC \\ 
                & D  & -0.89 & 0.02 & NC \\ 
                & E  & 0.90 & 0.01 & PC \\ 
                & F  & -0.99 & 0.00 & NC \\  
  
BL Lac	& A  & -0.78 & 0.00 & NC \\ 
        & B  & -0.63 & 0.00 & NC \\ 
        & C  & -0.80 & 0.02 & NC \\ 
        & D  & 0.54 & 0.01 & PC \\ 
        & E  & -0.94 & 0.00 & NC \\ 
        & F  & 0.90 & 0.00 & PC \\ 
        & G  & -0.80 & 0.02 & NC \\ 
        & H  & 0.90 & 0.00 & PC \\ 
        & I  & 0.80 & 0.02 & PC \\ 
        & J  & 0.81 & 0.03 & PC \\ 
        & K  & -0.77 & 0.04 & NC \\ 
        & L  & -0.77 & 0.02 & NC \\ 
        & M  & -0.74 & 0.04 & NC \\  
\hline 
\end{tabular}}
\end{table}

\begin{figure*}
\includegraphics[width=16cm, height=10cm]{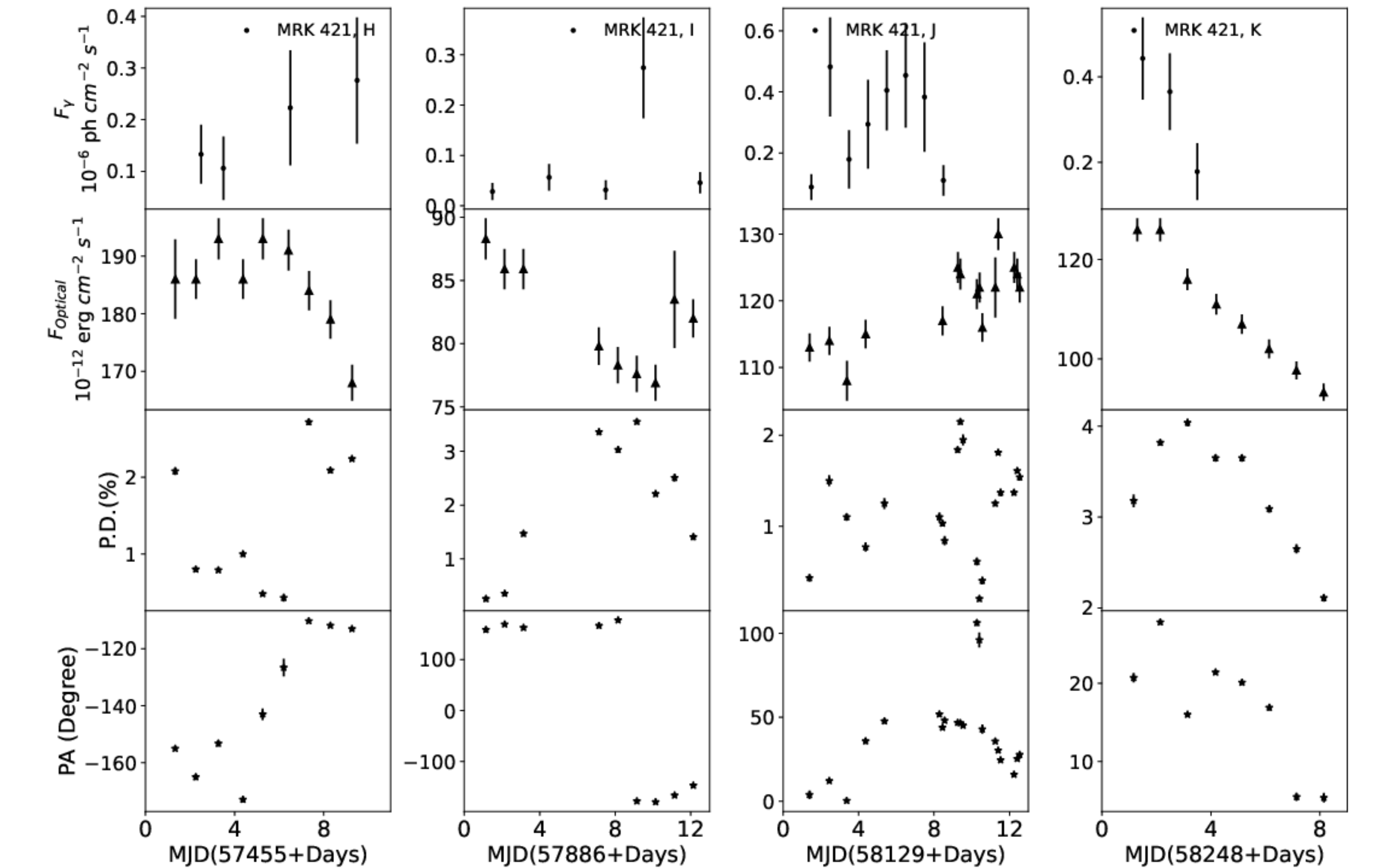}
\caption{\label{fig:epoch_lc1}Multi-wavelength light curves on short-term timescales. Each panel includes the name of the source and epoch. From top to bottom, the first panel represents the one-day binned $\gamma$-ray light curve, the second panel represents the light curve in the optical V-band, and the third and fourth panels represent the variation of optical polarization degree and position angle, respectively.}
\end{figure*}

\begin{figure*}
\centering
\includegraphics[width=18cm, height=18cm]{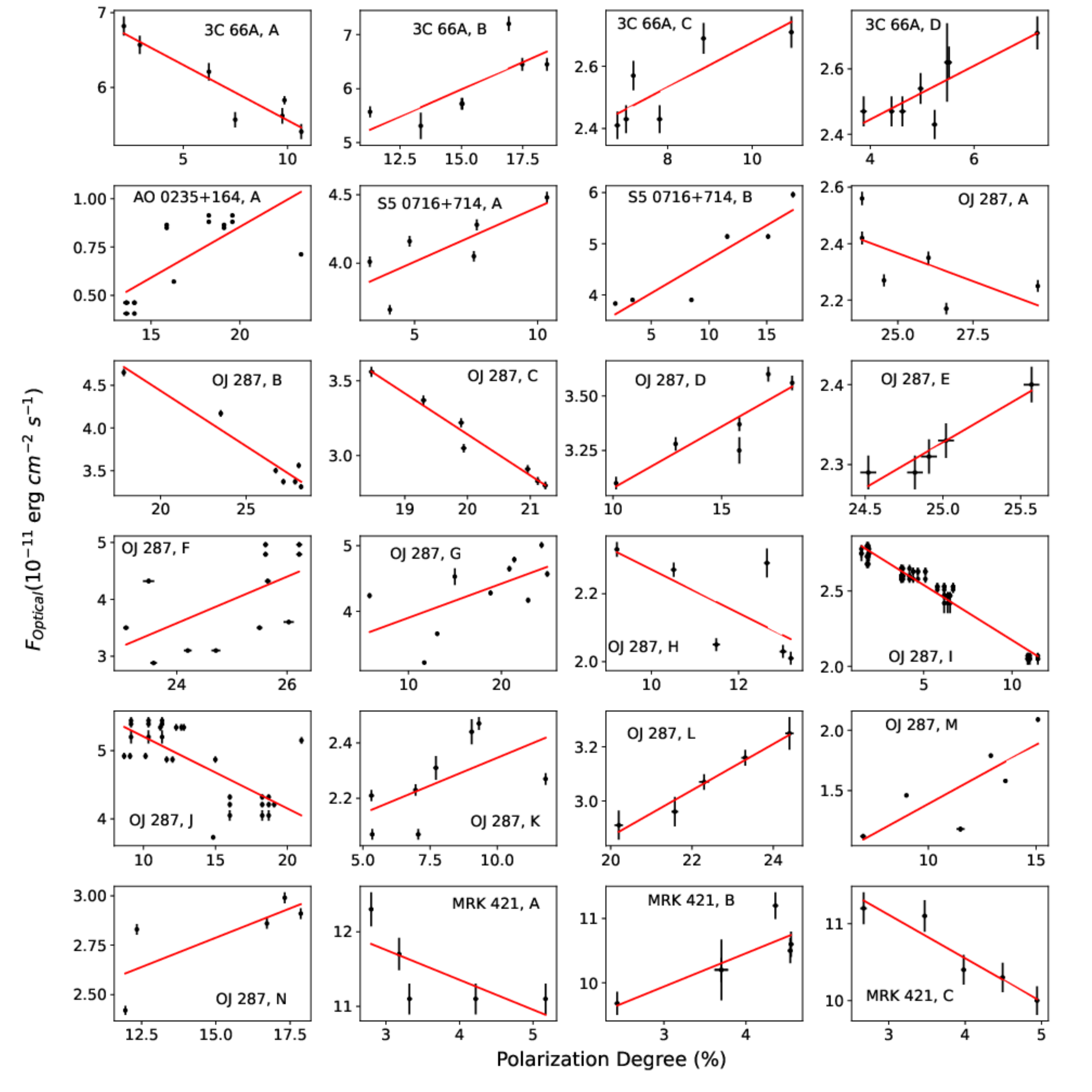}
\caption{\label{fig:opt_pol1} The correlation between the optical flux and the PD for 3C 66A, AO 0235+164, S5 0716+714, OJ 287 and MRK 421. Each panel includes the name of the source and the epoch, and the solid red line indicates the linear least square fit to the data.}
\end{figure*}

\section{ANALYSIS TECHNIQUES} \label{sec:analysis}
\subsection{Correlation between optical flux and polarization}
We looked for the correlation between optical flux and polarization degree on long-term as well as on short-term timescales. We used the same procedure as described in \cite{2022MNRAS.510.1809P}, to look for the correlation on different timescales. . 
\subsubsection{Correlation on long-term timescales}
The whole observation time period of $\sim$ 10 years was divided into 10 observing cycles and there is a gap of 4$-$6 months between each cycle. Table \ref{tab:LTV_stats} shows the time period of these observing cycles, the minimum, maximum, and average values of the $\gamma$-ray flux, optical flux and the polarization degree of each observing cycle for all the BL Lac sources.\\
To find out the correlation between optical flux and PD and  between optical flux and $\gamma$-ray flux we used the Spearman rank test. We considered the presence of significant positive correlation (PC) or negative correlation (NC) between optical flux and PD in a cycle, if the Spearman rank correlation coefficient R $>$ 0.5 or R $<$ -0.5, respectively, and the probability of null hypothesis P $<$ 0.05 (this corresponds to a 95\% level of confidence). The R and P values for each cycle are listed in Table. \ref{tab:LTV_stats}. The observing cycles which satisfy these criteria and the linear least-squares fit between optical flux and PD  and between $\gamma$-ray flux and optical flux are shown in Figures \ref{fig:pol_flx_cor1} and \ref{fig:opt_gamma_cor1}.
\subsubsection{Correlation on short-term timescales}
According to \cite{2018MNRAS.479.2037P} each observing cycle contains continuous monitoring for $\sim$ 10 days every month. We divided the light curves of each observation cycle into segments (epochs) for each source. So our short-term timescales contain a 10-days observation period. We estimated the optical flux variability amplitude for epochs that have at least 5 data points of optical flux, PD, and PA in the 10 day period.
\subsubsection{Optical flux variability amplitude}
The optical flux variability amplitude (fractional variability amplitude; $F_{var}$) is defined as \citep[e.g.][]{2003MNRAS.345.1271V, 2017ApJ...841..123P}
\begin{equation}
F_{var} = \sqrt{\frac{S^2 - \overline{\sigma_{err}^2}}{{\bar{x}^2}}}.
\end{equation}
In the above equation, $S^2$ is the sample variance of the light curve, $\overline{\sigma_{err}^2}$ is the mean square error. These are defined as:
\begin{equation}
S^{2} = \frac{1}{N-1}\sum_{i=1}^{N}(x_{i} - \bar{x})^{2}
\end{equation}
 and
\begin{equation}
\overline{\sigma^{2}_{err}} = \frac{1}{N}\sum_{i=1}^{N}{\sigma^{2}_{err,i}} \,
\end{equation}
The uncertainty in the $F_{var}$ is defined as:
\begin{equation}
err(F_{var}) = \sqrt{\Bigg(\sqrt{\frac{1}{2N}}\frac{\overline{\sigma_{err}^{2}}}{\bar{x}^{2}F_{var}}\Bigg)^{2} + \Bigg(\sqrt{\frac{\overline{\sigma_{err}^{2}}}{N}}\frac{1}{\bar{x}}\Bigg)^{2}}
\end{equation}
Using the same method as given in \cite{2022MNRAS.510.1809P}, we considered the source to be significantly variable in the epoch if $F_{var}$ $>$ 3$\times$err($F_{var}$). For the epochs where the source is significantly variable, we looked for a correlation between optical flux and PD. We considered that the optical flux and PD variations have significant positive or negative correlation if (i) the Spearman rank correlation coefficient is $>$ 0.5 or $<$ -0.5, respectively, and (ii) the probability of null hypothesis is $<$ 0.05. Only those epochs with a significant positive or negative correlation were considered for the further detailed study. The results of the statistical tests are given in Table \ref{tab:STV_stats}.  The results of the correlation analysis between optical flux and PD are given in Table \ref{tab:epoch_cor}.  The multiwavelength light curves of some epochs for the source MRK 421 are shown in Figure \ref{fig:epoch_lc1}. For the other epochs, the multiwavelength light curves are available in the electronic version.  The plots for the correlation between optical flux and PD for few sources are shown in Figure \ref{fig:opt_pol1}, while for the remaining sources, they are given in the electronic version. We give the values of fractional variability amplitude, F$_{var}$, and their corresponding errors for each of the epochs in Table \ref{tab:epoch_res}. 

\subsection{Optical variability timescale}
We calculated optical flux variability timescale $\tau_{var}$ for each epoch that satisfied our criteria as \citep{2013ApJ...773..147J}
\begin{equation}
    \tau_{var} = \frac{\Delta t}{ln(F_{2}/F_{1})}
\end{equation}
Here $F_{2}$ and $F_{1}$ are the optical flux values at time $t_{2}$ and $t_{1}$, respectively, and $\Delta$t is the time difference between any two consecutive flux measurement and defined as $\Delta$t = |$t_{2}$-$t_{1}$|. We took into account every possible pair of flux values that met the following conditions (i) $F_{2}$ $>$ $F_{1}$ and (ii) $F_{2}$ $-$ $F_{1}$ $>$ 3($\sigma_1$+$\sigma_2$)/2, where $\sigma_1$ and $\sigma_2$ are the uncertainties in the flux values $F_{1}$ and $F_{2}$, respectively. We considered the minimum value of $\tau_{var}$ as the variability timescale of the epoch. We estimated the error in $\tau_{var}$ using the error propagation method \citep{1992drea.book.....B}. The optical flux variability timescales and their corresponding errors for each source are given in Table \ref{tab:epoch_res}.

\subsection{Spectral Index}
We also calculated the spectral index for each epoch using the flux densities in V and R bands. The spectral index is defined as:
\begin{equation}\label{eq:alpha}
\alpha  = - \frac{ln\left(\frac{f_V}{f_R}\right)}{ln\left(\frac{\nu_V}{\nu_R}\right)},
\end{equation}
Where $f_{V}$ and $f_{R}$ are the flux densities ($f_{\nu}$ $\propto$ $\nu^{-\alpha}$) in the optical V and R bands, respectively and $\nu_{V}$ and $\nu_{R}$ are the effective frequencies in the V and R bands, respectively. We give the minimum, maximum and average values of spectral indexes for all the epochs in Table \ref{tab:epoch_res}.

\subsection{Determination of physical parameters using optical variability timescale}

We calculated the upper limit on the size of the emission region using the values of minimum variability timescale $\tau_{var}$ and Doppler factor $\delta$, as 
\begin{equation}
R \leq c \tau_{var} \frac{\delta}{1+z}
\end{equation}
They are given in Table \ref{tab:epoch_res}. We have adopted the values of the Doppler factor from \cite{2018ApJS..235...39C}, where the calculation of the Doppler factor is based on broadband spectral energy distribution (SED).\\
Given that the observed changes in the optical flux of BL Lac sources are due to synchrotron cooling, and that the observed minimum variability timescale must be greater than or equal to the lifetime of synchrotron electrons, $t_{syn}$, we calculated the magnetic field as:
\begin{equation}\label{eq:sync}
t_{syn} \approx 4.75 \times 10^2 \left(\frac{1+z}{\delta^o \nu_{GHz} B^3}\right)^{1/2} days,
\end{equation}
The estimated values of the magnetic field for the 11 BL lac sources are given in Table \ref{tab:epoch_res}.

\begin{table*}
\centering
\caption{\label{tab:epoch_res}Results of optical variability analysis and estimated physical parameters.}
\begin{tabular} {lcccccccccc}\hline
Blazar		& Epoch & $F_{var}(\%)$ & $\tau_{var}$   & \multicolumn{3}{c}{$\alpha$}  & R  & $B$  \\ 
        	&	&               &   (in days)             &   Min   & Max    &    Average              &    (in cm)     &   (in G)           \\ \hline 
3C 66A 	& A & 8.82$\pm$0.67   &   9.16$\pm$2.19 & 0.65 & 0.78 & 0.71 &   6.38$\times$$10^{17}$ & 0.06  \\ 
        & B & 10.69$\pm$0.89   &   8.61$\pm$2.04 & 0.77 & 0.88 & 0.81 &   5.99$\times$$10^{17}$ &  0.06 \\ 
        & C & 5.06$\pm$0.78   &   16.61$\pm$7.72 & 0.72 & 0.82 & 0.76 &   11.57$\times$$10^{17}$ &  0.04  \\ 
        & D & 3.07$\pm$0.97   &   17.02$\pm$7.52 & 0.77 & 0.90 & 0.82 &   11.85$\times$$10^{17}$ &  0.04   \\ 
		      
AO 0235$+$164   & A & 30.29$\pm$0.32   &   1.08$\pm$0.11 & 2.31 & 3.80 & 2.57 &  0.21$\times$$10^{17}$ &  0.36 \\ 
 
S5 0716$+$714   & A & 6.69$\pm$0.38   &   7.89$\pm$0.80 & 1.23 & 1.29  & 1.27  &  3.16$\times$$10^{17}$ &  0.08  \\ 
                & B & 19.22$\pm$0.38   &   3.59$\pm$0.17 & 1.22 & 1.28 & 1.24 &  1.44$\times$$10^{17}$ &  0.13  \\   
OJ 287	& A & 5.88$\pm$0.38   &   0.10$\pm$0.02 & 1.46 & 1.75 & 1.66 &   0.13$\times$$10^{17}$   &  0.93 \\ 
        & B & 14.15$\pm$0.38   &   3.80$\pm$0.19 & 1.40 & 1.42 & 1.41 &   5.09$\times$$10^{17}$   &  0.08 \\ 
        & C & 9.25$\pm$0.35   &   16.18$\pm$3.84 & 1.23 & 1.36 & 1.28 &   21.68$\times$$10^{17}$   &  0.03 \\ 
        & D & 5.60$\pm$0.46   &   17.57$\pm$4.06 & 1.30 & 1.37 & 1.34 &   23.54$\times$$10^{17}$   &  0.03 \\ 
        & E & 1.73$\pm$0.44   &   63.37$\pm$17.59 & 1.23 & 1.31 & 1.25 &   84.89$\times$$10^{17}$   & 0.01 \\ 
        & F & 20.24$\pm$0.31   &   4.31$\pm$1.61 & 1.25 & 1.41 & 1.34 &   5.77$\times$$10^{17}$   &  0.08 \\ 
        & G & 12.37$\pm$0.40   &   3.69$\pm$0.17 & 1.29 & 1.40 & 1.35 &   4.94$\times$$10^{17}$   &  0.08 \\ 
        & H & 6.74$\pm$0.47   &   17.75$\pm$3.30 & 1.29 & 1.42 & 1.35 &   23.78$\times$$10^{17}$   &  0.03 \\ 
        & I & 8.66$\pm$0.25   &   1.52$\pm$1.04 & 1.22 & 1.41 & 1.30 &   2.04$\times$$10^{17}$   &  0.15 \\ 
        & J & 12.58$\pm$0.33   &   2.72$\pm$1.24 & 1.36 & 1.52 & 1.44 &   3.64$\times$$10^{17}$   &  0.10 \\ 
        & K & 6.46$\pm$0.45   &   10.85$\pm$1.53 & 1.35 & 1.51 & 1.43 &   14.53$\times$$10^{17}$   &  0.04 \\ 
        & L & 4.29$\pm$0.71   &   27.27$\pm$15.41 & 1.29 & 1.36 & 1.34 &   36.53$\times$$10^{17}$   &  0.02 \\ 
        & M & 23.95$\pm$0.48   &   6.17$\pm$0.52 & 1.33 & 1.49 & 1.40 &   8.27$\times$$10^{17}$   &  0.06 \\ 
        & N & 7.87$\pm$0.41   &   21.30$\pm$6.23 & 1.40 & 1.42 & 1.41 &   28.53$\times$$10^{17}$   &  0.03 \\          
MRK 421 & A & 4.30$\pm$0.86   &   19.04$\pm$9.45 & 0.72 & 0.83 & 0.76 &   0.72$\times$$10^{17}$   &  0.09 \\ 
        & B & 4.19$\pm$1.11   &   20.33$\pm$10.12 & 0.70 & 0.72 & 0.71 &   0.77$\times$$10^{17}$   &  0.09 \\ 
        & C & 4.59$\pm$0.86   &   24.60$\pm$16.39 & 0.54 & 0.68 & 0.62 &   0.93$\times$$10^{17}$   &  0.08 \\
        & D & 7.44$\pm$0.38   &   1.46$\pm$0.59 & 0.49 & 0.72 & 0.59 &   0.06$\times$$10^{17}$   &  0.51 \\ 
        & E & 8.59$\pm$1.09   &   26.57$\pm$6.10 & 0.52 & 0.60 & 0.57 &   1.00$\times$$10^{17}$   &  0.07 \\ 
        & F & 5.87$\pm$0.77   &   48.71$\pm$15.49 & 0.54 & 0.62 & 0.59 &   1.84$\times$$10^{17}$   &  0.05 \\ 
        & G & 8.26$\pm$0.66   &   6.27$\pm$1.02 & 0.49 & 0.58 & 0.53 &   0.24$\times$$10^{17}$   & 0.19 \\ 
        & H & 3.66$\pm$0.77   &   15.28$\pm$6.29 & 0.54 & 0.62 & 0.59 &   0.58$\times$$10^{17}$   &  0.11 \\ 
        & I & 4.46$\pm$0.83   &   12.23$\pm$7.38 & 0.76 & 0.88 & 0.80 &  0.46$\times$$10^{17}$   &  0.12 \\ 
        & J & 4.38$\pm$0.57   &   2.47$\pm$1.60 & 0.58 & 0.67 & 0.62 &   0.09$\times$$10^{17}$   &  0.36 \\ 
        & K & 11.00$\pm$0.66   &   12.09$\pm$3.82 & 0.70 & 0.76 & 0.72 &   0.46$\times$$10^{17}$   &  0.12 \\ 
        & L & 9.72$\pm$0.76   &   18.04$\pm$8.44 & 0.71 & 0.94 & 0.87 &   0.68$\times$$10^{17}$   &  0.10 \\  
W Com   & A & 8.92$\pm$1.28   &   7.33$\pm$2.76 & 0.92 & 1.66 & 1.54 &   2.38$\times$$10^{17}$   &  0.08 \\
        & B & 4.96$\pm$1.54   &   7.10$\pm$2.75 & 1.30 & 1.49 & 1.38 &  2.30$\times$$10^{17}$   &  0.09 \\       
MRK 501	 & A & 2.42$\pm$0.61   &   52.20$\pm$35.59 & 1.66 & 1.72 & 1.67 &   3.01$\times$$10^{17}$   & 0.04 \\ 
         
1ES 1959$+$650  & A & 7.37$\pm$0.95   &   11.86$\pm$2.32 & 0.92 & 1.00 & 0.95 &  4.19$\times$$10^{17}$   &  0.06 \\ 
                & B & 3.61$\pm$0.74   &   9.47$\pm$2.35 & 0.94 & 1.06 & 1.00 &   3.35$\times$$10^{17}$   &  0.07 \\ 
                & C & 3.05$\pm$0.80   &   2.71$\pm$1.25 & 0.70 & 0.77 & 0.72 &   0.96$\times$$10^{17}$   &  0.16 \\ 
PKS 2155$-$304  & A & 4.13$\pm$0.33   &   9.99$\pm$1.29 & 0.85 & 0.92 & 0.89 &   2.58$\times$$10^{17}$   &  0.07 \\ 
                & B & 5.32$\pm$1.02   &   9.79$\pm$3.76 & 0.85 & 0.97 & 0.89 &   2.52$\times$$10^{17}$   &  0.08 \\ 
                & C & 2.99$\pm$0.38   &   14.99$\pm$2.91 & 0.84 & 0.92 & 0.87 &   3.86$\times$$10^{17}$   &  0.06 \\ 
                & D & 9.59$\pm$0.48   &   7.80$\pm$0.78 & 0.97 & 1.15 & 1.02 &   2.01$\times$$10^{17}$   &  0.09 \\ 
                & E & 2.57$\pm$0.48   &   18.39$\pm$4.36 & 0.86 & 0.92 & 0.90 &  4.74$\times$$10^{17}$   &  0.05 \\ 
                & F & 7.32$\pm$0.73   &   13.61$\pm$2.42 & 0.91 & 0.98 & 0.95 &   3.51$\times$$10^{17}$   &  0.06 \\ 
BL Lac	& A & 5.13$\pm$0.31   &   2.95$\pm$2.05 & 2.36 & 2.42 & 2.39 &   0.27$\times$$10^{17}$   & 0.24 \\ 
        & B & 7.01$\pm$0.25   &   2.61$\pm$0.75 & 2.36 & 2.48 & 2.41 &   0.24$\times$$10^{17}$   &  0.26 \\ 
        & C & 13.70$\pm$0.38   &   3.20$\pm$0.15 & 2.30 & 2.48 & 2.40 &   0.29$\times$$10^{17}$   &  0.22 \\ 
        & D & 30.73$\pm$0.30   &   0.88$\pm$0.08 & 2.19 & 2.36 & 2.30 &   0.08$\times$$10^{17}$   &  0.53 \\ 
        & E & 13.03$\pm$1.16   &   3.71$\pm$0.18 & 2.63 & 2.75 & 2.70 &   0.34$\times$$10^{17}$   &  0.20 \\ 
        & F & 25.36$\pm$0.40   &   4.24$\pm$0.23 & 2.39 & 3.45 & 2.59 &   0.39$\times$$10^{17}$   &  0.19 \\ 
        & G & 9.94$\pm$0.43   &   4.73$\pm$0.47 & 2.15 & 2.27 & 2.20 &   0.44$\times$$10^{17}$   &  0.17 \\ 
        & H & 10.59$\pm$0.47   &   3.83$\pm$0.19 & 2.20 & 3.44 & 2.41 &   0.35$\times$$10^{17}$   &  0.20 \\ 
        & I & 16.54$\pm$0.84   &   0.32$\pm$0.09 & 2.19 & 2.37 & 2.28 &   0.03$\times$$10^{17}$   &  1.04 \\ 
        & J & 12.54$\pm$0.35   &   10.89$\pm$0.70 & 2.30 & 2.47 & 2.41 &   1.00$\times$$10^{17}$   &  0.10 \\ 
        & K & 6.88$\pm$0.35   &   10.09$\pm$1.46 & 2.23 & 2.61 & 2.35 &   0.93$\times$$10^{17}$   &  0.10 \\ 
        & L & 13.08$\pm$0.53   &   6.09$\pm$0.76 & 2.13 & 2.39 & 2.29 &   0.56$\times$$10^{17}$   &  0.15 \\ 
        & M & 10.38$\pm$0.33   &   5.94$\pm$0.46 & 2.23 & 2.39 & 2.31 &   0.55$\times$$10^{17}$   &  0.15 \\  
 \hline
\end{tabular}
\end{table*}

\begin{figure*}\label{fig:q_u}
\centering
\includegraphics[width=18cm, height=14cm]{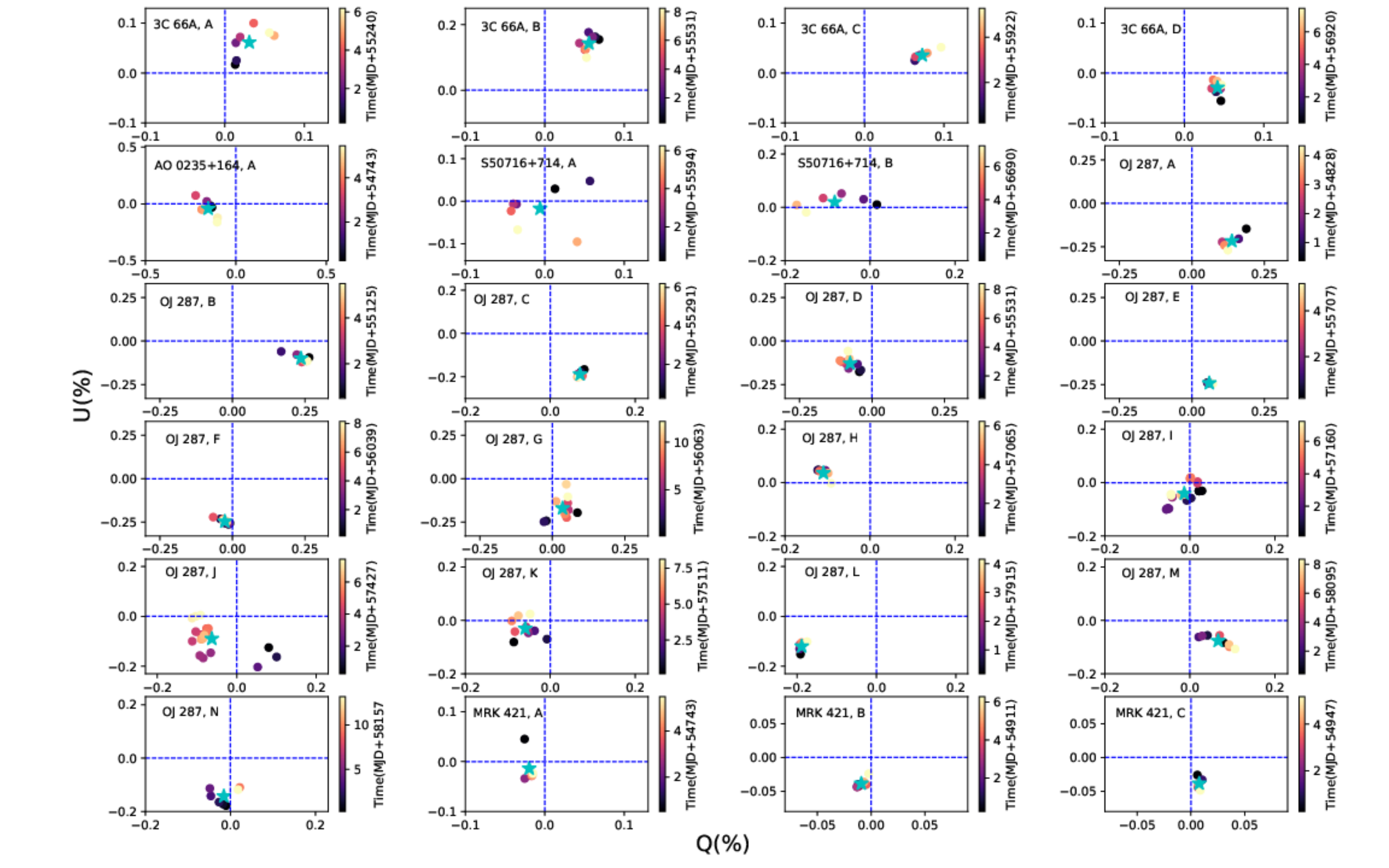}
\caption{\label{fig:q_u1} The relation between Stokes Q and Stokes U parameters for 3C 66A, AO 0235+164, S5 0716+714, OJ 287, and MRK 421. In each plot, the Q=0 and U=0 lines are shown as blue dashed lines, and the centroid is shown by a cyan star.}
\end{figure*}

\section{results}\label{sec:results}
We searched for correlation between optical flux and PD variations for the 11 BL Lac sources. We computed the flux variability amplitude and variability timescale for correlated and anti-correlated epochs of these sources. Using these parameters we further calculated the size of the emission regions and the magnetic field of the variable epochs. The following are the results of our analysis of our sample sources:
\subsection{3C 66A}
3C 66A, at a redshift z = 0.44 \citep{1978bllo.conf..176M} is classified as an intermediate synchrotron peaked BL Lac (IBL; \citealt{2015ApJ...810...14A}). This source was detected in high-energy $\gamma$-rays (HE; E $>$ 100 MeV) by $\it{Fermi}$-LAT \citep{2016ApJS..222....5A} and also in very high-energy $\gamma$-ray regime (VHE; E $>$ 100 GeV) by VERITAS and MAGIC telescopes \citep{2009ApJ...693L.104A, 2011ApJ...726...58A}. It was observed to show intraday variability in the infrared and optical bands \citep{1997A&A...318..331D, 2012MNRAS.425.3002G, 2017MNRAS.469.2457L, 2018AJ....155...90F, 2017MNRAS.469.2305K}. In October 2008, this source showed a $\gamma$-ray outburst that was observed by $\it{Fermi}$-LAT and VERITAS and found to be correlated with an optical outburst in the R-band \citep{2011ApJ...726...43A}. The optical light curves of this source over a period of 3 years revealed evidence of quasi-periodic variability \citep{2020MNRAS.492.5524O}. Variability characteristics in the GeV band was reported by \cite{2013ApJ...767..103V}. They found variation on timescales as short as 1.5 days.\\
This source was observed from MJD 55088 to MJD 58306 by the Steward Observatory. During this period the optical V-band flux ranged from 1.53 $\times$ $10^{-11}$ erg $cm^{-2}$ $s^{-1}$ to 8.50 $\times$ $10^{-11}$ erg $cm^{-2}$ $s^{-1}$ and the degree of polarization ranged from 0.21\% to 19.68\%. For all of the observing cycles, we found no significant correlation between optical flux and PD and between optical flux and $\gamma$-ray flux (Table \ref{tab:LTV_stats}). On short-term timescales, we found significant positive correlation in three out of four epochs, and significant negative correlation in one epoch (see Table \ref{tab:epoch_cor}). We calculated flux variability amplitude on shorter time-scale, which varied from 3.07\% to 10.69\%. On short-term timescales during epochs A and B the $\gamma$-ray flux was nearly constant within error bars, whereas optical flux and PD varied significantly. The $\gamma$-ray flux was not detected during epochs C and D, but there were variations in the optical flux and PD. Using optical flux variability timescales we found the values of R $\sim$ (5.99$-$11.85)$\times$$10^{17}$ cm and B $\sim$ (0.04$-$0.06) G. 

\subsection{AO 0235+164}
AO 0235+164 at a redshift of z = 0.94 \citep{1987ApJ...318..577C}, was one of the first objects identified as a BL Lac object based on its featureless optical spectrum \citep{1975ApJ...201..275S}. It is a low synchrotron peaked BL Lac (LBL; \citealt{2010ApJ...716...30A}). This source has shown variation over the whole of electromagnetic spectrum \citep{1976Natur.260..752L, 1976Natur.260..754R,  2009A&A...507..769R, 2008ATel.1784....1F, 2010ApJ...716...30A}. The radio light curve for this source showed quasi-periodic oscillations \citep{2001A&A...377..396R, 2021MNRAS.501.5997T}. This source has been observed to display a very high and variable degree of polarization from radio to optical bands \citep{1979ApJ...229L...1L, 1982MNRAS.198....1I}. The positive correlation between optical flux and PD was observed by \cite{2008ApJ...672...40H} during the outburst in December 2006. \cite{2011PASJ...63..489S} reported a positive correlation between optical flux and PD during the 2008-2009 activity state. In numerous studies, the variation between optical and $\gamma$-ray flares was found to be well correlated for this source \citep{2012ApJ...751..159A, 2018ARep...62..103H, 2021MNRAS.504.1772R}, implying that the optical and $\gamma$-ray emitting locations are in the same region of the jet.\\
AO 0235+164 was observed by the Steward Observatory from MJD 54743 to MJD 58161. It showed optical flux changes during this observing period, with V-band flux ranging from (0.03$-$1.32) $\times$ $10^{-11}$ erg $cm^{-2}$ $s^{-1}$. The minimum and maximum values of PD were found to be 0.58\% and 39.79\% respectively. We identified a positive correlation between optical flux and PD for one (C4) out of ten observing cycles during long-term observation, whereas no significant correlation was found between optical flux and PD in the remaining cycles. The brightness of the $\gamma$-ray flux and optical flux changed slightly during cycle C4, but there were significant changes in PD and PA. We also found positive correlation between optical flux and $\gamma$-ray flux during the cycles C1 and C8. On short-term timescale we found one epoch where optical flux and PD are positively correlated. During the same time span, \cite{2011PASJ...63..489S} also found a positive correlation between optical flux and PD on short-term timescale. We calculated the flux variability amplitude to be 30\% for this epoch. The $\gamma$-ray and optical fluxes increased during this epoch and there was also a change in PD. During this epoch, PA displayed a mixed pattern of behaviour, first decreasing and then increasing. We derived values of R $\sim$ 0.21$\times$$10^{17}$ cm and B $\sim$ 0.36 G.\\

\subsection{S5 0716+714}
S5 0716+714 is an IBL \citep{2010ApJ...715..429A, 2010ApJ...716...30A} located at redshift z = 0.31 \citep{2008A&A...487L..29N}. This source has been thoroughly studied for short-term and long-term optical variability \citep{1997A&A...327...61G, 2003A&A...402..151R, 2006MNRAS.366.1337S, 2009MNRAS.399.1357S, 2019ApJ...880..155L, 2019ApJ...884...92X} and known to show highly variable nature across a wide frequency range \citep{1996AJ....111.2187W, 2013AdSpR..51.2358R, 2020ApJ...904...67G}. The microvariability events studied for this source in optical band were found to be characterized by high degree of polarization ($>$ 30\%) \citep{2016ApJ...831...92B}. This source is also known to exhibit QPOs in optical band on timescales of 25$-$73 minutes and 15 minutes \citep{2009ApJ...690..216G, 2010ApJ...719L.153R}. During the optical outburst in April 2008, the $360^{\circ}$ rotation in the electric vector position angle was recorded in this source by \cite{2008ATel.1502....1L}. The detection of $\gamma$-ray activity by the AGILE satellite in September and October 2007 showed the correlation behaviour with optical flux \citep{2008A&A...489L..37C}.\\
The Steward Observatory observed this source from MJD 54743 to MJD 58158. During this observing period of $\sim$ 9.5 years, the optical V-band flux changed from 2.07 $\times$ $10^{-11}$ erg $cm^{-2}$ $s^{-1}$ to 30.41 $\times$ $10^{-11}$ erg $cm^{-2}$ $s^{-1}$ and PD changed from 0.49\% to 27.96\%. On long-term timescales, we identified a positive correlation between optical flux and PD in one (C9) out of 10 observing cycles and positive correlation between optical flux and $\gamma$-ray flux during cycle C7, while in the remaining cycles no significant correlation was observed. Large changes in $\gamma$-ray flux and moderate changes in optical were seen during cycle C9. PD changed dramatically during this cycle, but PA showed moderate changes. On short-term timescales, we found two epochs with significant positive correlation between optical flux and PD. The flux variability amplitude ranged from 6.7\% to 19.2\% on short-term timescale. During these two epochs $\gamma$-ray flux was not detected and there was an increase in optical flux and PD. On short-term timescales, variations in PA were seen during these two epochs but no abrupt change was noticed. We estimated the value of R $\sim$ (1.44$-$3.16)$\times$$10^{17}$ cm and B $\sim$ (0.08$-$0.13) G.\\

\subsection{OJ 287}
OJ 287 is a LBL source \citep{2015ApJ...810...14A}, at redshift z = 0.306 \citep{2010A&A...516A..60N}. \cite{1988ApJ...325..628S} discovered periodicity behaviour using long-term optical light curve, and found that the double-peaked outburst is repeated in every $\sim$ 11.65 years. This is attributed to the secondary super massive black hole striking the accretion disc around the first super massive black hole in the binary super massive black hole system. This source has been immensely studied in optical band for its flux variability behaviour \citep{1971ApL.....9..151A, 2011MNRAS.413.2157R, 2011MNRAS.416..101G, 2017ApJ...835..275R, 2017MNRAS.465.4423G, 2017ApJ...844...32P, 2019AJ....157...95G, 2021RAA....21..138Y} as well as in other bands \citep{2011ApJ...726L..13A, 2018MNRAS.473.1145K, 2020MNRAS.498L..35K, 2020MNRAS.499..653K, 2020MNRAS.498.3578S}. The correlation analysis between optical flux and polarization has been carried out in many studies. The correlated behaviour between the optical flux (R-band) and PD was observed by \cite{2014MNRAS.439..639B} in the period 2012 November$-$2013 April and during this period variation in position angle was also observed that was corresponding to rotation of the polarization plane of 5.80 deg $d^{-1}$. During the 2016 outburst this source was monitored by \cite{2017ApJ...835..275R} and they found a significant anti-correlation between optical flux and polarization on intranight timescale. The counterclockwise looping was observed between flux and PD variations on 2014 February 20 by \cite{2020MNRAS.492.1295P}. The optical and $\gamma$-ray flux variations in this source were found to be correlated \citep{2021MNRAS.504.1772R}. The possible quasi-periodic oscillation of $\sim$ 314 days in the 9.5 year $\gamma$-ray light curve was studied by \cite{2020MNRAS.499..653K}.\\
The Steward Observatory observed this source for a period of $\sim$ 10 years, from MJD 54743 to MJD 58292. The optical V-band flux varied from 0.74 $\times$ $10^{-11}$ erg $cm^{-2}$ $s^{-1}$ to 7.72 $\times$ $10^{-11}$ erg $cm^{-2}$ $s^{-1}$ and PD varied from 0.73\% to 36.54\% during this observing period. On long-term timescales we identified two out of ten observing cycles, where optical flux and PD are positively correlated in cycle C4 and negatively correlated in cycle C7. In the remaining 8 cycles we didn't observe any significant correlation between optical flux and PD. During cycle C4, both the $\gamma$-ray flux and optical flux changed, with short-term flares superimposed over long-term variations. Optical flux changes appeared to reflect changes in $\gamma$-ray flux. Changes in the PD seemed to follow the optical flux during this cycle, but PA remained relatively constant. During cycle C7, variations in the $\gamma$-ray flux were reported, and short-term optical flares are superimposed on long-term optical changes. During this cycle, there were changes in the PD, as well as an abrupt change in the PA of $254^{\circ}$. We also noticed positive correlation between optical flux and $\gamma$-ray flux during cycle C2 and negative correlation during cycle C4 on long-term timescales. On short-term timescales, we found 14 epochs where optical flux and PD were significantly correlated. Optical flux and PD showed a positive correlation in 8 epochs and a negative correlation in 6 epochs. For these epochs the flux variability amplitude ranged from 1.7\% to 24\%. During epochs A, D, E, F, H, I, K, L, M and N, the $\gamma$-ray flux was not detected, while in other epochs changes in the $\gamma$-ray flux have been observed. In any of the epochs, no abrupt changes in PA were seen. We derived R $\sim$ (0.13$-$84.89)$\times$$10^{17}$ cm and B $\sim$ (0.01$-$0.93) G using optical flux variability timescales.\\

\subsection{MRK 421}
MRK 421 is one of the closest (redshift z = 0.031; \citealt{1991rc3..book.....D}) and brightest objects in the very high energy (VHE) extragalactic sky. It is a HBL source \cite{2010ApJ...716...30A} and showed frequent flaring activity in the GeV and TeV emission. It is the first extragalactic source which is detected in TeV energy range \citep{1992Natur.358..477P}. This source has been extensively studied across the wide range of electromagnetic spectrum (i.e. from radio to $\gamma$-rays) due to its unremarkable variability nature \citep{2004ApJ...601..759T, 2005ApJ...630..130B, 2011ApJ...736..131A, 2015A&A...578A..22A, 2017ApJ...841..123P, 2018MNRAS.480.4873A, 2020MNRAS.494.3432G, 2021A&A...647A..88A}. It displayed intranight variability in R-band flux and PD variations during April 2013 \citep{2017ApJS..232....7F}. In this flaring period the source showed the varied behaviour of correlation and anti-correlation between optical and TeV $\gamma$-ray fluxes.\\
MRK 421 was observed by the Steward Observatory for a period of $\sim$ 10 years from MJD 54744 to MJD 58306. During this observing period the optical V-band flux ranged from 5.9 $\times$ $10^{-11}$ erg $cm^{-2}$ $s^{-1}$ to 34.2 $\times$ $10^{-11}$ erg $cm^{-2}$ $s^{-1}$ and PD ranged from 0.05\% to 12.4\%. On long-term timescales, we looked for a correlation between optical flux and PD for ten observation cycles and didn't find any significant correlation in any of them. We found positive correlation between optical flux and $\gamma$-ray flux during cycle C5. On short-term timescales we found 12 epochs, where significant correlation between optical flux and PD was observed. We found a positive correlation between optical flux and PD in 6 of the 12 epochs, while a negative correlation was found in the other 6 epochs. During these 12 epochs the flux variability amplitude varied from 3.7\% to 11\%. The $\gamma$-ray flux was not detected during epoch A and it was nearly constant within the error bar during epoch B. In other 10 epochs the $\gamma$-ray flux varied moderately. During epoch A, we noted a sudden shift in PA of $60^{\circ}$, and during epoch I, we noticed an abrupt change of $355^{\circ}$. PA did not show any abrupt changes during other epochs. We found R $\sim$ (0.06$-$1.84)$\times$$10^{17}$ cm and B $\sim$ (0.05$-$0.51) G.\\

\subsection{W com}
W Comae is an IBL source \citep{2006A&A...445..441N}, discovered in 1971 \citep{1971Natur.231..515B}. This source is located at redshift z = 0.103 \citep{1985ApJ...292..614W}. This source was detected by EGRET at $\gamma$-ray energies in the energy range of 100 MeV to 10 GeV \citep{1999ApJS..123...79H}. W Comae was the first IBL which was detected at very high energies by VERITAS \citep{2008ApJ...684L..73A}. This source has shown optical flux variability on diverse timescales \citep{1995PASP..107..863S,  2012MNRAS.425.3002G}. This source had a remarkable optical outburst (R mag = 12.2) in April-May 1998, which was detected by \cite{1999A&A...342L..49M}. The time lag analysis between optical and $\gamma$-ray for this source was studied by \cite{2018MNRAS.480.5517L} who found the lag of $\sim$ $8^{+7.8}_{-7.8}$ days with the $\gamma$-ray flux leading the optical flux variations.\\
This source was observed by Steward Observatory from MJD 54768 to MJD 58306. During the $\sim$ 10 years the optical V-band flux changed in the range (0.66$-$4.12) $\times$ $10^{-11}$ erg $cm^{-2}$ $s^{-1}$ and the PD changed from 0.81\% to 26.7\%. In two (C1 and C9) out of ten observing cycles, we found a significant positive correlation between optical flux and PD, but in the other cycles, we found no significant correlation between optical flux and PD. During cycle C1, changes in the $\gamma$-ray flux were noticed. The changes in PD seemed to follow the changes in the optical flux, but the PA remained nearly constant during cycle C1. During cycle C9, the $\gamma$-ray flux was not detected but drastic change in the optical flux was noticed. During this cycle, there were significant changes in the PD and PA. We didn't find any significant correlation between optical flux and $\gamma$-ray flux on long-term timescales. On short-term timescales, we identified two epochs with significant correlations between optical flux and PD. Optical flux and polarization are anti-correlated in epoch A while they are positively correlated in epoch B. On short-term timescales the flux variability amplitude varied in the range 5\%$-$9\%. During epochs A and B the $\gamma$-ray flux was not detected, whereas the optical flux and PD showed variations. During both epochs, there were no abrupt changes in the PA. We found R $\sim$ (2.30$-$2.38)$\times$$10^{17}$ cm and B $\sim$ (0.08$-$0.09) G.\\

\subsection{MRK 501}
MRK 501, at a redshift of z = 0.034 \citep{1975ApJ...198..261U} is a HBL \citep{2010ApJS..188..405A} and one of the brightest extragalactic objects in the TeV $\gamma$-ray sky. It is the second object which was detected in the very high energy (VHE; $>$ 300 GeV) $\gamma$-rays \citep{1996ApJ...456L..83Q}  This source has been extensively studied over the wide range of the frequencies which include radio, optical, X-ray, and $\gamma$-ray using several observational facilities \citep{1999ApJ...518..693Q, 2001ApJ...546..898A, 2004A&A...422..103M, 2015ApJ...812...65F, 2017ApJ...841..123P, 2017A&A...603A..31A, 2020MNRAS.492.2261S}. This source showed intranight optical variability on May 20, 2015 \citep{2017ApJ...844...32P}. During the observation period from 2010 to 2015, it showed optical flux variation in the V, R, and I bands on diverse timescales \citep{2016ApJS..222...24X}. During a 4.5 months multiwavelength campaign from March 15 to August 1, 2009, in the $\gamma$-ray range from 0.1 GeV to 20 TeV this source was monitored by $\it{Fermi}$-LAT, MAGIC, VERITAS and Whipple 10m telescopes \citep{2017A&A...603A..31A}. The $\gamma$-ray flux varied significantly in the VHE regime, whereas the optical flux remained nearly constant. Quasi-periodic oscillation of 330-days was reported in 5 of the 7 cycles in $\gamma$-ray light curve observed by $\it{Fermi}$-LAT over a ten-year period from August 2008 to August 2018 \citep{2019MNRAS.487.3990B}.\\
Steward Observatory monitored this source for $\sim$ 10 years (MJD 54745-58306). During this monitoring period the optical V-band flux varied from 5.08 $\times$ $10^{-11}$ erg $cm^{-2}$ $s^{-1}$ to 6.82 $\times$ $10^{-11}$ erg $cm^{-2}$ $s^{-1}$ and PD varied from 0.07\% to 5.93\%. On long-term timescales, we found 3 out of 10 observing cycles, where significant correlation between the optical flux and PD was observed. During these 3 cycles C1, C2 and C6, we found a positive correlation between optical flux and PD. In the remaining 7 cycles the optical flux and PD were not significantly correlated. During these three cycles the $\gamma$-ray flux and optical flux were found to be variable. During cycle C1, a large change in the PD and PA was observed and PD also varied significantly during cycles C2 and C6. During cycle C2, PA showed small variations but during cycle C6, it was almost constant. We also found negative correlation between optical flux and $\gamma$-ray flux during cycle C1. On short-term timescales, we found one epoch where optical flux and PD were significantly positively correlated. The flux variability amplitude of this epoch is estimated to be 2.4\%. During this epoch, the $\gamma$-ray flux varied slightly, and no sudden changes in the PA were seen. Using optical flux variability timescales, we calculated the values of R $\sim$ 3.01$\times$$10^{17}$ cm and B $\sim$ 0.04 G.\\

\subsection{1ES 1959+650}
1ES 1959+650 is a HBL, located at a redshift z = 0.048 \citep{1996ApJS..104..251P}. This source is well studied in a wide range of electromagnetic wavebands \citep{1991ApJS...75.1011G, 2003A&A...406L...9A, 2004ApJ...601..151K, 2014ApJ...797...89A,  2017ApJ...841..123P, 2017ApJ...846..158K, 2018A&A...611A..44P, 2019MNRAS.490..124M, 2020A&A...638A..14M}. Since 2015, the source has been active across several energy bands, most notably in optical \citep{2016ATel.9070....1B}, X-rays \citep{2016ATel.9205....1K}, and $\gamma$-rays, monitored by the the MAGIC, $\it{Fermi}$-LAT and VERITAS collaborations \citep{2016ATel.9010....1B, 2016ATel.9148....1B}. It has also been studied for short-term and long-term flux variations \citep{2012MNRAS.420.3147G, 2016A&A...593A..98L, 2017MNRAS.469.1682Z}. The study of the polarization position angle variability for this source has been carried out by \cite{2016A&A...596A..78H} among the sample of TeV and non-TeV blazars using data from the RoboPol blazar monitoring program and the Nordic Optical Telescope. \cite{2018MNRAS.480.5517L} found the time lag of $\sim$ $5.5^{+6.1}_{-6.1}$ days between optical and $\gamma$-ray flux variations with the $\gamma$-ray flux leading the optical flux variations.\\
This source was observed by Steward Observatory from MJD 54745 to MJD 58293. During this 10 years monitoring period the optical V-band flux varied in the range (1.95$-$7.70) $\times$ $10^{-11}$ erg $cm^{-2}$ $s^{-1}$ and PD varied between 0.37\%$-$8.53\%. In one (C2) out of ten observing cycles, we identified a positive correlation between optical flux and PD, while in one cycle (C4), we found a negative correlation. In the remaining 8 cycles we didn't observe any significant correlation between optical flux and PD. During cycle C2, the $\gamma$-ray flux was nearly constant, whereas short-term variations in the optical flux were found to be superimposed upon long-term changes. Large changes in the PD and moderate changes in the PA were observed during this cycle. During cycle C4, the $\gamma$-ray flux did not vary, but considerable changes in the optical flux and PD were observed. PA also varied during this cycle.  We found no significant correlation between optical flux and $\gamma$-ray flux in any cycle. On short-term timescales we found three epochs where optical flux and PD were significantly correlated. There was a positive correlation between optical flux and PD during epochs A and C, but a negative correlation between optical flux and PD during epoch B. The estimated value of flux variability amplitude varied from 3.1\% to 7.4\% on a short-term timescales. During epochs A and B, the $\gamma$-ray flux was not detected, and during epoch C, it showed a variable nature. For this source, we determined the values of R $\sim$ (0.96$-$4.19)$\times$$10^{17}$ cm and B $\sim$ (0.06$-$0.16) G.\\

\subsection{PKS 2155$-$304}
PKS 2155$-$304 is a HBL source at a redshift of 0.116 \citep{1984ApJ...278L.103B} and very well known object in the southern hemisphere. It has been found to be variable across wavelengths \citep[e.g.][and references therein]{1998A&A...333L...5G, 2007ApJ...664L..71A, 2008A&A...484L..35F, 2017MNRAS.466.3309R, 2017ApJ...841..123P, 2019MNRAS.484..749C, 2021ApJ...909..103Z}. The $\it{Fermi}$-LAT reported variability of this source on month-like time scles in the GeV energy range \citep{2015ApJS..218...23A}. This source showed varied behaviour of correlation and anti-correlation between optical flux and position angle during the optical outburst in 2010 \citep{2020MNRAS.495.2162P}. The first evidence of quasi-periodic oscillations in the polarized flux was found in PKS 2155$-$304 \citep{2016MNRAS.462L..80P}. According to this study the source displayed periodic component at T $\sim$ 13 min and T $\sim$ 30 min during intraday polarization monitoring from July 25 to July 27, 2009. QPOs were also found in the optical R-band and $\gamma$-ray light curves with timescales of 0.87 year \citep{2014RAA....14..933Z} and 1.74 year \citep{2017ApJ...835..260Z} respectively. The correlated behaviour between optical and GeV $\gamma$-ray flux variations was seen by \cite{2021MNRAS.504.1772R}. An optical flare without $\gamma$-ray counterpart both in GeV and very high energy bands was noticed in 2016 by \cite{2019hepr.confE..27W}.\\
PKS 2155$-$304 was observed by Steward Observatory from MJD 54745 to MJD 58306. During this 10 years monitoring period the optical V-band flux ranged from 3.0 $\times$ $10^{-11}$ erg $cm^{-2}$ $s^{-1}$ to 23 $\times$ $10^{-11}$ erg $cm^{-2}$ $s^{-1}$ and PD ranged from 0.17\% to 19.1\%. On long-term timescales we identified significant correlation between optical flux and PD in 4 out of 10 observing cycle. All four cycles C6, C7, C9 and C10 exhibited positive correlation between optical flux and PD. We didn't find any significant correlation in the remaining 6 cycles. During cycles C6, C7, C9 and C10, the $\gamma$-ray flux was found to be variable and it showed the largest flare in the 10-year period during cycle C6. During these four cycles, the optical flux and PD were variable, and during cycles C9 and C10, short-term optical flares were superimposed on long-term optical changes. The PA was found to vary during cycles C7 and C 10. On long-term timescales, we also found positive correlation between optical flux and $\gamma$-ray flux during cycles C1, C3 and C10. On short-term timescales we found 6 epochs where the optical flux and PD are found to be significantly correlated. During epochs A, B, C and E the correlation between optical flux and PD was positive, whereas the correlation was negative during epochs D and F. The flux variability amplitude varied between 2.6\%$-$10\% on short-term timescales. The $\gamma$-ray flux was not detected considerably during epochs A, B, D, and F, and it showed random variable behaviour during epochs C and E. We didn't notice any abrupt change in the PA during any of the epochs. We estimated R $\sim$ (2.01$-$4.74)$\times$$10^{17}$ cm and B $\sim$ (0.05$-$0.09) G.\\

\subsection{BL Lac}
BL Lac is located at redshift z = 0.069 \citep{1977ApJ...212L..47M}. It is an IBL \citep{2011ApJ...743..171A}. It is well known for its variable behaviour in all energy bands \citep{2008Natur.452..966M, 2013MNRAS.436.1530R, 2015MNRAS.450..541A, 2015MNRAS.452.4263G, 2016ApJ...816...53W, 2019A&A...623A.175M, 2020ApJ...900..137W}. The variability timescale of this source was found to be $\sim$ 30 minutes at optical frequency observed by TESS (Transiting Exoplanet Survey Satellite) from September 12 to October 6, 2019 \citep{2020ApJ...900..137W}. The variation in PD in the optical R-band was also seen in this study, with a timescale of $\sim$ 5 hours, where the PD changed by a factor of two, and changes in the optical flux and the PD were not found to be correlated. The intraday optical variability was also reported by \cite{2017MNRAS.469.3588M} in the optical B, V, R and I bands for 13 nights between 2012 and 2016. \cite{2017A&A...600A.132S} found periodic variability $\sim$ 680 days in the optical R-band and $\gamma$-ray light curve over an 8 years period from 2008 to 2016. In the long-term study ($\sim$ 6.5 years) of GeV bright blazars \cite{2016ApJ...833...77I} found the positive correlation with no time lag between optical and $\gamma$-ray flux variations.\\
During the 10 years monitoring period by Steward observatory, the optical V-band flux varied in the range (1.6$-$23.0) $\times$ $10^{-11}$ erg $cm^{-2}$ $s^{-1}$ and the PD varied between 0.62\%$-$26.08\%. On long-term timescales we found one cycle (C3) out of 10 observing cycles, which showed significant negative correlation between optical flux and PD. During other cycles we found no significant correlation between optical flux and PD. During cycle C3, the $\gamma$-ray flux varied and changes in the optical flux appeared to follow the changes in the $\gamma$-ray flux. Large changes have been reported in the PD and PA during this cycle. We found positive correlation between optical flux and $\gamma$-ray flux on long-term timescales during cycles C3 and C7. On short-term timescales, we identified 13 epochs with significant positive correlation between optical flux and PD in 5 epochs (D, F, H, I and J) and significant negative correlation in 8 epochs (A, B, C, E, G, K, L and M). The flux variability amplitude varied in the range (5.13\%$-$30.7\%) on short-term timescales. During the epochs A, B, C, E, F, G, and J, either no or less $\gamma$-ray flux was detected but it varied during other epochs. We didn't observe any dramatic change in the PA during any of the epochs. For this source, we found the values of R $\sim$ (0.03$-$1.00)$\times$$10^{17}$ cm and B $\sim$ (0.10$-$1.04) G.\\

\subsection{1ES 2344+514}
1ES 2344+514 first discovered by $\it{Einstein}$ Slew Survey \citep{1992ApJS...80..257E} in the energy range 0.2-4 KeV, is at a redshift of z = 0.044 \cite{1996ApJS..104..251P}. In the VHE $\gamma$-ray this source was first detected in 1995 by the $\it{Whipple}$ 10m telescope during an intense flare \citep{1998ApJ...501..616C}. In 1996 this source was at its brightest state in X-ray band \citep{2000MNRAS.317..743G} and the synchrotron peak frequency ($\nu_{s}$) had shifted by a factor of 30 or more. Because of this shift in the $\nu_{s}$, this source is included in the Extreme high-frequency BL Lacs (EHBL) class, whose synchrotron peak of the broad-band spectral energy distribution was at $\nu_{s}$ $\geq$ $10^{17}$ Hz. After this unprecedented event many multiwavelength campaigns were organized to study this source \citep{2011ApJ...738..169A, 2013A&A...556A..67A, 2020MNRAS.496.3912M}. This source showed intraday variability in the optical R-band \citep{2010Ap&SS.327...35M}. \cite{2020MNRAS.496.1430P} detected the signature of significant IDV in the optical R-band. \cite{2018MNRAS.480.5517L} found no significant correlation between the optical and $\gamma$-ray flux variations for this source.\\ 
This source was observed by Steward observatory for a time period of $\sim$ 10 years from MJD 54743 to MJD 58282. During this monitoring period, the optical V-band flux varied between (1.6$-$2.6) $\times$ $10^{-11}$ erg $cm^{-2}$ $s^{-1}$ and PD varied between 0.45\%$-$5.63\%. On both long- and short-term timescales, we found no significant correlation between optical flux and PD.


\begin{table*}
\centering
\caption{\label{tab:res_sum}Summary of the results}
\resizebox{1.0\textwidth} {!}{ 
\begin{tabular} {lccccccccccc}\hline
Blazar		& \multicolumn{4}{c}{Long-term timescale} & \multicolumn{3}{c}{Short-term timescale} &  R  & $B$  \\ \cmidrule[0.03cm](r){2-5}\cmidrule[0.03cm](r){6-8}
        	& No. of observing cycles &  PC   &  NC &  No Corr. & No. of epochs & PC & NC                          &    ( $\times$ 10$^{17}$ cm)     &   (in G)  \\ \hline
3C 66A		& 9 & 0 & 0 & 9 &  4	& 3 & 1 &   5.99$-$11.85 &  0.04$-$0.06 \\
AO 0235$+$164  & 10 & 1 & 0 & 9 &  1	& 1 & 0   &   0.21 &  0.36 \\
S5 0716$+$714  & 10 & 1 & 0 & 9 &  2	& 2 & 0 &   1.44$-$3.16 &  0.08$-$0.13      \\  
OJ 287		& 10 & 1 & 1 & 8 &  14	& 8 & 6  &   0.13$-$84.89 &  0.01$-$0.93  \\
MRK 421	& 10 & 0 & 0 & 10 &  12	& 6 & 6  &  0.06$-$1.84 &   0.05$-$0.51       \\
W Com 	& 10 & 2 & 0 & 8 &  2	& 1 & 1  &   2.30$-$2.38 &  0.08$-$0.09        \\
MRK 501	& 10  & 3 & 0 & 7 &  1	& 1 & 0 &    3.01 &  0.04       \\
1ES 1959+650	& 10 & 1 & 1 & 8 &  3	& 2 & 1  &  0.96$-$4.19 &  0.06$-$0.16        \\
PKS 2155$-$304  & 10 & 4 & 0 & 6 & 6 & 4 & 2 &  2.01$-$4.74 &  0.05$-$0.09\\
BL Lac  & 10 & 0 & 1 & 9 & 13 & 5 & 8 &  0.03$-$1.00 &  0.10$-$1.04\\
1ES 2344+514  & 10 & 0 & 0 & 10 & 0 & 0 & 0 & --&--\\
 \hline
\end{tabular}}
\end{table*}

\begin{table*}
\centering
\caption{\label{tab:vlbi_sum}Comparison of optical polarization data and 15 GHz VLBI polarization data.}
\begin{tabular} {lccccccc}\hline
Blazar	&	& \multicolumn{3}{c}{Optical} & \multicolumn{3}{c}{VLBI}    \\ \cmidrule[0.03cm](r){3-5}\cmidrule[0.03cm](r){6-8}
        & Epoch	& Period &  PD(\%)   &  PA (degree) &  Epoch & PD(\%) & PA (degree) \\ \hline 
OJ 287 & B & 55125-55130 & 25.63$\pm$0.02 & -11.30$\pm$0.04 & 55129 & 7.5  &   169\\
       & G & 56063-56075 & 17.88$\pm$0.02 & -36.79$\pm$0.05 & 56071 & 8.9  &   140\\
 \hline
\end{tabular}
\end{table*}


\section{DISCUSSION AND CONCLUSION} \label{sec:diss}
 We searched for a systematic correlation between optical flux and PD for 11 BL Lac sources on long-term as well as short-term timescales. To find the correlation between optical flux and PD we used $\sim$ 10 years of Steward observatory archival data. On long-term timescales, we found significant positive correlation (R $>$ 0.5 \& P $<$ 0.05) in 13 cycles, significant negative correlation (R $<$ $-$0.5 \& P $<$ 0.05) in 3 cycles and no correlation in 93 observing cycles. On short-term timescales, we found a total of 58 epochs with significant correlation. We found significant positive correlation (R $>$ 0.5 \& P $<$ 0.05) in 33 of the 58 epochs and significant negative correlation (R $<$ $-$0.5 \& P $<$ 0.05) in 25 of the 58 epochs. We also looked for a correlation between optical flux and $\gamma$-ray flux on long-term timescales. We found significant positive correlation (R $>$ 0.5 \& P $<$ 0.05) in 10 cycles and significant negative correlation (R $<$ $-$0.5 \& P $<$ 0.05) in 2 cycles. In the remaining 89\% of the observing cycles we did not find any significant correlation. We give the summary of our correlation analysis in Table \ref{tab:res_sum}.\\
The results of our correlation analysis of optical flux and PD in BL Lac sources show a wide range of behaviors. We found positive as well as a negative connection between optical flux and polarization degree. Extrinsic scenarios in which shocks propagating along relativistic jets and slightly deviating from linearity also known as "swinging jet model", can explain the correlation (or anti-correlation) between optical flux and PD \citep{1992A&A...259..109G}. The observed positive correlation between optical flux and PD can also be explained by intrinsic scenario shock-in-jet model \citep{2008ApJ...672...40H,2013ApJS..206...11S}. Shocks in the jets cause the relativistic electrons to be injected into the emission area, resulting in a stable spectral shape and short timescale variability in the optical band. In addition, shocks are responsible for the alignment of the magnetic field. So, the shock-in-jet scenario leads to a positive correlation between optical flux and polarization degree.\\
We also observed anti-correlated behaviour between optical flux and PD on long-term as well as short-term timescales. \cite{2008Natur.452..966M} developed a scenario in which the radiation source is made up of two or more emission regions, one of which is a global jet region and the others are local emission regions. Here the local emission occurs from highly polarized shocked "clumps" travelling inside the jet and is characterized by short-term variability because of the small emission regions. Similar approach is also adopted in \cite{2014ApJ...780...87M}, where the Turbulent, Extreme Multi-zone (TEMZ) model can explain the anti-correlation between optical flux and PD. In local emission region, an increase in optical flux is caused by newly produced polarized components, whereas multiple randomly oriented polarized components with similar strengths and varying position angles produce partial cancellations \citep{2002A&A...385...55H}. The degree of polarization decreases as a result of it. We plotted the observed Stokes parameters Q and U for each epoch of all the sources to find the presence of a stable polarized component and other multiple components. These are shown in Figure \ref{fig:q_u1}, for other epochs these plots are available in the electronic version. In the Q vs U plot, the divergence from the centroid (0,0) indicates the presence of a stable polarized component \citep{2009A&A...503..103B}. Additionally, we searched for any specific patterns that might be present between Q and U during both the positive and negative correlations of optical flux and PD. We noticed that the relationship between Q and U is entirely arbitrary and unrelated to either the positive or negative correlation between optical flux and PD. \\
We also examined to see whether any of the VLBI observations overlapping during any of the epochs we studied as VLBI observations could provide parsec-scale resolution. We found 2 VLBI observations at 15 GHz \cite{2018ApJS..234...12L}, overlapping with the 2 epochs of the source OJ 287. We give the value of the VLBI observation for these two epochs in Table \ref{tab:vlbi_sum}. The VLBI PD values do not match the optical PD values and are lower than the optical PD values. This is because VLBI measurements are collected on a single date, whereas optical data are averaged over a 10-day period that includes the VLBI observation epoch. Moreover, radio emission emanates from larger regions of the jets, the larger emission region may lead to the decrease in the PD due to different orientation of the magnetic field vectors within the VLBI beam. In contrast, we found that the PA values at optical and VLBI observations are same (after a $180^{\circ}$ correction). This suggests that the jet's trajectory from the optical to the millimetre scale is rather straight \citep{2006ChJAS...6a.247J}. We identified a negative and positive correlation between optical flux and PD during the epochs B and G of OJ 287, respectively. The observed flux variation could be attributed to this new emission region if the increased flux variations coincide with the formation of a new VLBI knot (a new emission region). A correlation between flux and PD is expected if the magnetic field in this new blob of emission is aligned with the large scale magnetic field. An anti-correlation between flux and PD can be expected if the direction of the magnetic field of the new emitting blob is either chaotic or misaligned with the large scale magnetic field.\\
To check for variations in the polarization degree, we also adopted a method based on changes in the spectral index of the emitting electron in the jets \citep{2018ApJ...858...80R}. For the power law distribution of the relativistic electrons (dN/dE $\propto$ $E^{-p}$; p = 2$\alpha$+1), the maximum degree of polarization is defined by
\begin{equation}
    PD = \frac{\alpha+1}{\alpha+5/3}
\end{equation}
Here $\alpha$ is the spectral index. For each BL Lac sources, we calculated spectral index using equation \ref{eq:alpha}. The minimum, maximum and average values of the spectral index for each epochs of the BL Lac sources are given in Table \ref{tab:epoch_res}. The variance in the PD was determined using the minimum and maximum values of the spectral index for each source. From our calculations, we found maximum change in PD is 2\% for 3C 66A, 5\% for AO 0235+164, 1\% for S5 0716+714, 2\% for OJ 287, 2\% for MRK 421, 6\% for W Com, 0.4\% for MRK 501, 1\% for 1ES 1959+650, 2\% for PKS 2155$-$304 and 4\% for BL Lac. But the maximum change in PD from observation is 8.5\% for 3C 66A, 9.8\% for AO 0235+164, 15.4\% for S5 0716+714, 19.1\% for OJ 287, 6.1\% for MRK 421, 4.7\% for W Com, 1.8\% for MRK 501, 2.76\% for 1ES 1959+650, 10.4\% for PKS 2155$-$304, and 12.5\% for BL Lac. The maximum change in PD due to a change in the spectral index is much smaller than the observed PD variations. It implies that the observed variations in PD in this study cannot be explained exclusively by changes in the power-law spectral index of the emitting electrons within the jet.\\
We also noticed a sudden decline of $355^{\circ}$ in PA during one epoch I of the source MRK 421. During this abrupt change the $\gamma$-ray flux and optical flux both changed, but the polarization degree decreased (see Figure \ref{fig:epoch_lc1}). Optical polarization angle swings have been found to be linked to multiwavelength flares \citep{2016MNRAS.463.3365A, 2018MNRAS.474.1296B}. According to \cite{2018ApJ...862L..25Z}, relativistic magnetic reconnection could be driving their multi-wavelength flares and these multiwavelength flares are also accompanied by changes in optical polarization angles. As observed by \cite{2020ApJ...901..149Z}, large rotations in the polarization angle are important indicators of magnetic reconnection between antiparallel magnetic field lines and during large angle swings, this antiparallel magnetic field causes a decrease in PD. These PA swings are most commonly associated with changes in the $\gamma$-ray flux, implying that they are physically connected to $\gamma$-ray activity \citep{2015MNRAS.453.1669B, 2018MNRAS.474.1296B}.\\
We also observed both correlation and anti-correlation between optical flux and $\gamma$-ray flux. The observed correlation between these two can be consistent with leptonic models in which optical and $\gamma$-ray emission are produced by the same population of electrons via synchrotron and inverse Compton processes, respectively \cite{2021MNRAS.504.1772R}. And anti-correlation between optical flux and $\gamma$-ray flux can be explained by a change in the magnetic field, without a change in the total number of emitting electrons or the Doppler factor of the emitting region \citep{2013ApJ...763L..11C,2020MNRAS.498.5128R}.\\

\section{Summary} \label{sec:sum}
For 11 $\gamma$-ray bright BL Lacs, we used $\sim$ 10 years of optical photometric and polarimetric data from Steward Observatory, as well as $\gamma$-ray data from $\it{Fermi}$-LAT, and looked for (i) correlation between optical V-band flux and PD on long-term as well as short-term timescales, and (ii) correlation between optical V-band flux and $\gamma$-ray flux on long-term timescales. The conclusions of our study are summarised below.
\begin{enumerate}
    \item On long-term timescales, all 11 BL Lac sources were found to be variable in the $\gamma$-ray, optical, and PD. We found 16 observing cycles with significant correlation (P $<$ 0.05) between optical flux and PD on long-term timescales, with significant positive correlation in 13 cycles and significant negative correlation in 3 cycles. In 93 observing cycles, we noticed no significant correlation. 
    \item On short-term timescales, we found a total of 58 epochs showing a significant correlation (P $<$ 0.05) between optical flux and PD. We found significant positive correlation in 33 out of 58 epochs and significant negative correlation in 25 out of 58 epochs.  
    \item On long-term timescales, we found a significant correlation (P $<$ 0.05) between optical flux and $\gamma$-ray flux, with positive correlation identified in 10 observing cycles and negative correlation found in 2 observing cycles. In the remaining 89\% of the observing cycles, there was no significant correlation between optical flux and $\gamma$-ray flux.
    \item On short-term timescales, the optical flux variability amplitude of the sources varied from 1.73\% to 30.73\%, the variability timescale varied from 0.10 days to 63 days and the resulting magnetic fields varied in the range 0.01 to 1.04 G.
\end{enumerate}
\section*{Acknowledgements}
 We thank the anonymous referee for his/her valuable suggestions, which helped to make the manuscript better. This research used Fermi data from the NASA Goddard Space Flight Center's Fermi Science Support Center (FSSC). We are thankful for photometry and polarimetry from Paul Smith’s monitoring program at the Steward Observatory, which is supported by Fermi Guest Investigator grants NNX08AW56G, NNX09AU10G, and NNX12AO93G. This research has made use of data from the MOJAVE database that is maintained by the MOJAVE team (Lister et al. 2018). We also acknowledge the use of High Performance Computing Facility (Nova cluster) at Indian Institute of Astrophysics.

\section*{Data Availability}
The $\gamma$-ray data and optical data used in this work are publicly available from the Fermi-LAT\footnote{\url{https://fermi.gsfc.nasa.gov/ssc/data/access/}} and Steward Observatory\footnote{\url{http://james.as.arizona.edu/~psmith/Fermi}} data archives. It may also be made available to anyone on reasonable request to the lead author.

\bibliographystyle{mnras}
\bibliography{master}

\end{document}